\documentclass[letterpaper]{article} 
\usepackage{aaai24}  
\usepackage{times}  
\usepackage{helvet}  
\usepackage{courier}  
\usepackage[hyphens]{url}  
\usepackage{graphicx} 
\usepackage{natbib}  
\usepackage{caption} 
\usepackage{algorithm}
\usepackage{algorithmic}
\usepackage{amsmath}
\usepackage{amssymb}

\usepackage{multirow}

\usepackage{newfloat}
\usepackage{listings}
\DeclareCaptionStyle{ruled}{labelfont=normalfont,labelsep=colon,strut=off} 
\floatstyle{ruled}
\newfloat{listing}{tb}{lst}{}
\floatname{listing}{Listing}
\title{Percentile Risk-Constrained Budget Pacing for Guaranteed Display Advertising in Online Optimization}
\author{
    Liang Dai\textsuperscript{\rm 1},
    Kejie Lyu\textsuperscript{\rm 1},
    Chengcheng Zhang\textsuperscript{\rm 1},
    Guangming Zhao\textsuperscript{\rm 2},
    Zhonglin Zu\textsuperscript{\rm 1},
    Liang Wang\textsuperscript{\rm 2},
    Bo Zheng\textsuperscript{\rm 2}
}
\affiliations{
    \textsuperscript{\rm 1}Alibaba Group, Hangzhou, China\\
    \textsuperscript{\rm 2}Alibaba Group, Beijing, China\\
    \{dailiang.dl, lvkejie.lkj, xuelun.zcc, lambert.zgm, zhonglin.zuzl, liangbo.wl, bozheng\}@alibaba-inc.com\\
}

\begin{document}

\maketitle

\begin{abstract}
    Guaranteed display (GD) advertising is a critical component of advertising since it provides publishers with stable revenue and enables advertisers to target specific audiences with guaranteed impressions. However, smooth pacing control for online ad delivery presents a challenge due to significant budget disparities, user arrival distribution drift, and dynamic change between supply and demand. This paper presents robust risk-constrained pacing (RCPacing) that utilizes Lagrangian dual multipliers to fine-tune probabilistic throttling through monotonic mapping functions within the percentile space of impression performance distribution. RCPacing combines distribution drift resilience and compatibility with guaranteed allocation mechanism, enabling us to provide near-optimal online services. We also show that RCPacing achieves $O(\sqrt{T})$ dynamic regret where $T$ is the length of the horizon. RCPacing's effectiveness is validated through offline evaluations and online A/B testing conducted on Taobao brand advertising platform.
\end{abstract}

\section{Introduction}\label{section:Introduction}

According to a report by the Internet Advertising Bureau, online display advertising generated a remarkable revenue of \$63.5 billion in 2022, demonstrating a substantial year-over-year increase of 12.0\% \cite{ref:iab}. Ad exposures in display markets are sold through both guaranteed and non-guaranteed (like real-time bidding or RTB) selling channels. Within the guaranteed display (GD) selling channel, an advertiser (demand side) and publisher (supply side) negotiate a fixed price (cost-per-mille or CPM) for ad placement, including details such as when, where, and how the ad campaigns will be displayed. These contractual arrangements guarantee the delivery of a specified number of impressions that meet specific targeting criteria during a specified period.

In addition to contractual agreements, advertisers usually expect their ad campaigns to be delivered smoothly and steadily during the purchased period for various reasons, including making campaign performance as good as possible, reaching a wider audience, increasing the display ratio of the target audience, and maintaining stable online viewership for live streaming events. However, smooth and robust pacing control for hundreds or thousands of GD advertisements on a brand advertising platform that deals with billions of daily requests is a challenging task. To summarize, the main challenges are as follows:

\begin{itemize}
    \item Significant differences among campaigns: the guaranteed daily impressions range from thousands to millions, the targeted audience size also vary greatly. Moreover, different campaigns have different optimization goals, such as click-through rate (CTR) or conversion rate (CVR).
    \item Drastic changes in traffic environment: these changes include significant fluctuations in overall traffic, dynamic shifts in the distribution of user arrival over time, and the impact of other campaigns going online or offline.
\end{itemize}

The existing smooth pacing techniques have primarily focused on RTB ads \cite{nuara2022online, liu2020effective}, which is incompatible with GD allocation. Although some research has considered the smoothness or representativeness in online optimal allocation of GD ads, it is often not optimized and evaluated as a separate key metric. In this paper, we consider smooth pacing for GD ads from the perspective of a publisher. Our contributions can be summarized as follows:

\begin{itemize}
    \item We introduce a novel framework called RCPacing, which employs Lagrangian dual multipliers to adjust probabilistic throttling based on monotonic functions within the percentile space, allowing us to effectively manage risk and ensure optimal ad delivery performance.
    \item We also show that RCPacing attains regret of order $O(\sqrt{T})$ when the length of the horizon $T$ and the initial number of resources are scaled proportionally.
    \item As there exists a tradeoff between smooth and optimal allocation in online matching problems, RCPacing offers flexible control over this balance.
    \item We implement RCPacing in our online display advertising system and conduct extensive online/offline experimental evaluations. The results demonstrate that RCPacing is highly effective in improving both the performance and smoothness of online delivery for GD campaigns.
\end{itemize}
\section{Related Work}\label{section:Related Work}


In the past few years, the allocation of GD advertising has received significant attention from researchers \cite{wu2021impression,wang2022conflux}. It is typically modeled as an online matching problem, intending to achieve the maximum match between impressions and contracts\cite{chen2012ad}. While the primary objective is to provide each advertiser with a predetermined number of display opportunities, it is also necessary to consider the smoothness of budget consumption. Some researchers include a fixed representative term in objectives \cite{fang2019large,dai2023fairness,bharadwaj2012shale}, which aims to minimize the deviation between the allocation probability and its corresponding supply-demand ratio for each contract. However, the representative term is fixed without consideration of dynamic adjustment.


Another research direction is to achieve budget pacing through feedback control, which can be further categorized into bid modification \cite{mehta2007adwords, zhou2021primal} and probabilistic throttling \cite{agarwal2014budget, xu2015smart, lee2013Real}. Bid modification influences the budget spending of an ad by adjusting its bidding price. Mehta et al. \cite{mehta2007adwords} modify the bid by multiplying it with a value that reflects the proportion of unused budget, and the impression would be allocated to the ad with the highest modified bid. Balseiro et al. \cite{balseiro2020dual} and Zhou et al. \cite{zhou2021primal} utilize a dual multiplier to form a virtual bid, which is consistently updated based on the variance between the actual and expected budget consumption. These methods adhere to the same principle of decreasing an ad's bid when the budget is being spent too rapidly. However, both the dramatic change in the bid win-rate curve and bid landscape make it challenging to control the budget through bid modification. 

On the other hand, probabilistic throttling methods decouple spending control from bid calculation, they directly impact the participating likelihood based on its budget consumption speed. Agarwal et al. \cite{agarwal2014budget} set a global pass-through rate (PTR), which is decreased when the budget consumption speed exceeds the expectation, and increased when the consumption speed falls below. Although this method demonstrates good budget control capability, it heavily relies on the accuracy of traffic forecasting. 

To further consider performance optimization while achieving budget control, Xu et al. \cite{xu2015smart} group requests with similar response rates (e.g. CTR) together and share a PTR among them. When the PTRs need to be adjusted, the average response rate of each group determines the priority of that group. While effective in budget control, relying solely on PTR regulation is insufficient to ensure guaranteed display for GD allocation.
\section{Preliminaries}\label{section: Preliminaries}

\subsection{Problem Formulation}
We formulate the GD allocation problem as the following optimization problem:
\begin{equation}
    \begin{split}
        \max_{x^{(t)}\in\mathcal{X}}\sum_{t=0}^{T-1}f_t(x^{(t)})=\sum_{t=0}^{T-1}{v^{(t)}}^{\top}x^{(t)} \\
        \text{s.t.} \sum_{t=0}^{T-1}x^{(t)}\le B
    \end{split}
\end{equation}
where $x^{(t)} \in \mathcal{X} \subseteq \mathbb{R}^M$ is the one-hot decision vector at time $t \in \left[1,\; T\right]$, $M$ is the total number of campaigns and the impression arrived at time $t$ would be allocated to the $j$-th campaign if the $j$-th component of $x^{(t)}$ is 1, $v^{(t)} \in R^M$ denotes the impression quality between the impression and the campaigns, $f_t\left(x^{(t)}\right)={v^{(t)}}^{\top}x^{(t)} \in \mathbb{R}$ is the revenue obtained at time $t$, $B \in \mathbb{R}^M$ is the positive budget vector which represents the campaign budgets. 

Following Balseiro et al \cite{balseiro2020dual}., we define the offline dual problem as
\begin{equation}
    \label{eq:dual_problem}
    \begin{aligned}
        \min_{\alpha \ge 0}D\left(\alpha\right)&=\sum_{i=1}^np_if_i^*\left(\alpha\right)+\alpha^{\top}\rho \\
        &=\sum_{i=1}^np_i\max_{x \in \mathcal{X}}\left\{v^{(i)\top}x-\alpha^{\top}x\right\}+\alpha^{\top}\rho
    \end{aligned}
\end{equation}
where $f_i^*\left(\alpha\right):=\max_{x \in \mathcal{X}}\left\{f_i\left(x\right)-\alpha^{\top}x\right\}$ is the conjugate function of $f_i\left(x\right)$ (restricted in $\mathcal{X}$ ), $p_i$ is the probability that the $i\text{-th}$ impression has a quality vector of $v^{(i)}$ , n is the total number of impressions, $\rho=B/T$ is the average budget for each time period, $\alpha$ is the dual variable, and the $j$-th element of $\alpha$ (denoted as $\alpha_j$ ) reflects the additional revenue generated by allowing one unit of resources to be added to the $j$-th campaign's budget.

\subsection{Dual Mirror Descent Algorithm}

Our method is built upon the Dual Mirror Descent (DMD) algorithm \cite{balseiro2020dual}, which addresses the general online allocation problem with budget constraint. At time $t$, DMD filters out campaigns that have exhausted their budget and assigns the request to the campaign that offers the highest premium among the remaining campaigns (equation \ref{eq: alloc_dmd}). The dual variable is then updated according to the online mirror descent (equation \ref{eq: update_dmd}). More details about DMD is given in Algorithm \ref{alg: DMD}.

\begin{algorithm}
\caption{Dual Mirror Descent Algorithm}
\label{alg: DMD}
\begin{algorithmic}[1]
\REQUIRE Time period $T$, remaining resources $B^{(0)}=T\rho$, reference function $h\left(\cdot\right):\mathbb{R}^M \rightarrow \mathbb{R}$, and step-size $\eta$.
\STATE $\alpha^{(0)} = 0$
\FOR{$t=0$ \textbf{to} $T-1$}
\STATE Receive $v^{(t)} \sim \mathcal{P}$
\STATE Make the decision $\tilde{x}^{(t)}$ and update the remaining resources $B^{(t+1)}$, where
\begin{equation}
\label{eq: alloc_dmd}
    \tilde{x}_{j}^{(t)} = \left\{
        \begin{aligned}
            1, & \; \text{if } j = \mathop{\arg\max}\limits_{B_{j}^{(t)} \ge 1}\left\{v_{j}^{(t)}-\alpha_j^{(t)}\right\}\\
            0,& \; \text{otherwise.} \\
        \end{aligned}
    \right.
\end{equation}
\begin{equation}
    B^{(t+1)}=B^{(t)}-x^{(t)}
\end{equation}
\STATE Obtain a stochastic sub-gradient of $D\left(\alpha^{(t)}\right)$
\begin{equation}
    \tilde{g}^{(t)} :=-\tilde{x}^{(t)}+\rho
\end{equation}
\STATE Update the dual variable by mirror descent
\begin{equation}\label{eq: update_dmd}
\begin{aligned}
\alpha^{(t+1)}=\mathop{\arg\min}\limits_{\alpha \ge 0}\left<\tilde{g}^{(t)}, \alpha \right> + \frac{1}{\eta}V_h\left(\alpha, \alpha^{(t)}\right)& \\
\text{where }V_h\left(\alpha, \alpha^{(t)}\right) = h\left(\alpha\right)-h\left(\alpha^{(t)}\right)-&\\
\left<\nabla h\left(\alpha\right),\alpha-\alpha^{(t)}\right>&
\end{aligned}
\end{equation}

\ENDFOR
\end{algorithmic}
\end{algorithm}
\section{Motivation}\label{section:motivation}
\subsection{Assumptions}
In this paper, we adopt the following assumptions:
\begin{itemize}
\item \textbf{The small bids assumption}. Each impression has only one slot available for displaying ads, which is significantly lower than the demand for campaigns and the supply of publishers \cite{2012Online}.
\item \textbf{The known IID assumption}. The known Independent and Identically Distributed (IID) assumption implies that impressions arrive online according to a known probability distribution with repetition \cite{huang2021online}, which is a realistic assumption in our problem.
\end{itemize}

\subsection{Motivation}
\begin{figure}[h]
\centering
\includegraphics[width=0.49\textwidth]{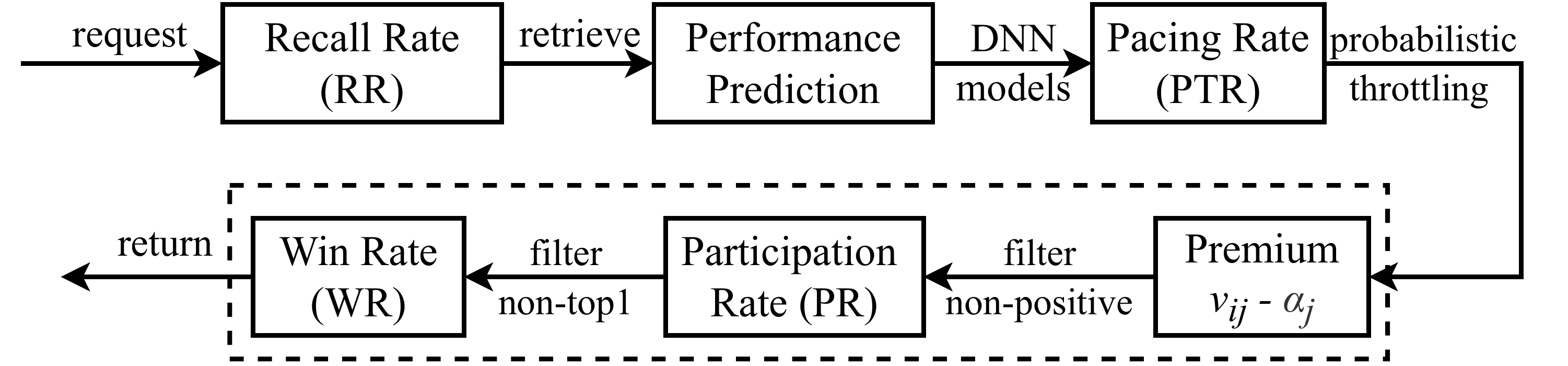}
\caption{Online process for GD campaign $j$.}
\vspace{-2ex}
\label{fig:primal_dual_online}
\end{figure}

Upon receiving a user's online request, the ad engine retrieves GD campaigns with recall rate (RR) that meet the targeting criteria and employs real-time prediction models such as deep neural networks (DNN) to estimate performance scores for each campaign \cite{zhou2019deep,gao2021advances}. The decision-maker then determines whether and which campaign to display based on the scores. The detailed processing flow is illustrated in figure\ref{fig:primal_dual_online}. The campaign $j$ passes the pacing control module with the pass-through rate (PTR) before the calculation of the ``price premium". It will only win out if it has the highest positive price compared to all other campaigns. The positive ratio of price and the win-out ratio are referred to as the participation ratio (PR) and win rate (WR). Without loss of generality, we use CTR as performance score in the following paragraph. The cost for campaign $j$ can be denoted as:
\begin{equation} 
\mathbb{E}\left[Cost_{j}\right] = \mathbb{E}\left[RR_{j}\right]\sum  PTR_{j}PR_{j} WR_{j}
\end{equation}

\begin{figure}[h]
\centering
\includegraphics[width=0.45\textwidth]{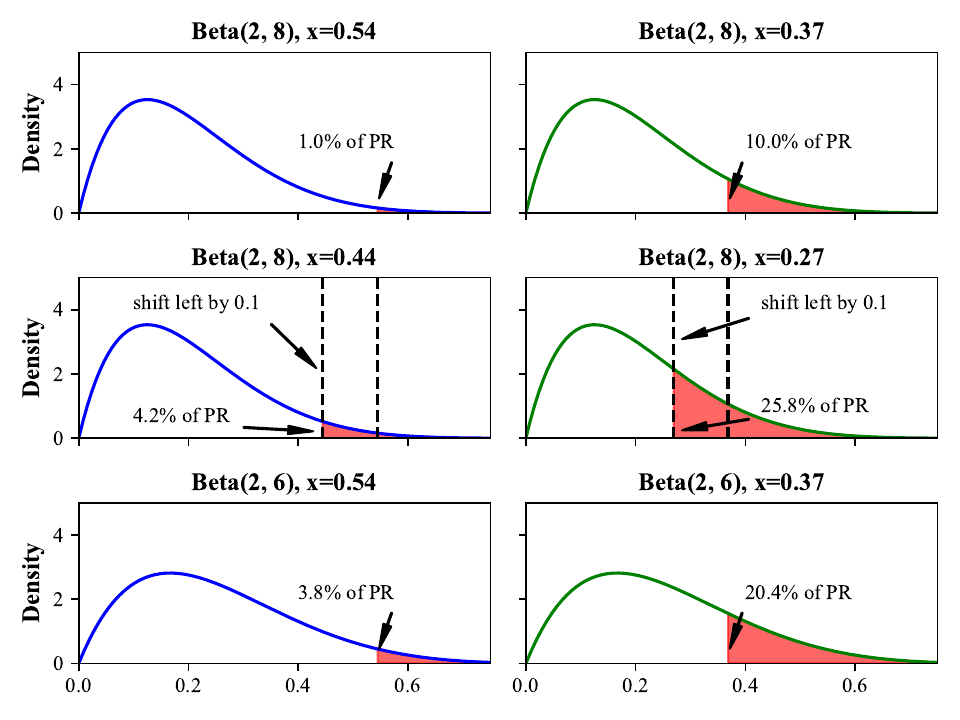}
  \vspace{-2ex}
\caption{Different changes of PR when adjusting for dual variables or distribution drift with the same magnitude.}
\vspace{-2ex}
\label{fig:beta_dist}
\end{figure}

It is worth noting that the primary risk GD campaigns is over-spend of the budget because it is irreversible once the budget is over-spent. Apart from PTR, the delivery of GD campaign is primarily determined by PR and WR. Let's consider two GD campaigns Ad1 and Ad2, with identical impression budgets, performance distributions, and similar competitive environments, but with different supply amounts (Ad1 $>$ Ad2). When the online allocation reaches a stable state, the dual variable of Ad1 is located at a higher percentile than Ad2 in performance distribution.
\begin{itemize}
\item \textbf{Risk analysis under stable conditions}: A higher percentile indicates a greater potential available traffic for Ad1, which makes it more vulnerable to over-spending. Moreover, Ad1 is more challenging to initialize dual variables because a higher percentile implies higher uncertainty especially before the start of delivery.
\item \textbf{Risk analysis under dynamic conditions}: Higher percentile results in a smaller bid price, making the campaign more susceptible to over-acceleration if other campaigns suddenly go offline. Moreover, as shown in the figure \ref{fig:beta_dist}, the PR of Ad1 generates greater fluctuations if the dual variable shifts the same distance, and is more sensitive to distribution drift in user arrival or switching of online prediction models, which can be deduced by the proof of Theorem 2 and Theorem 3 in appendix.
\end{itemize}

Based on the above risk analysis, RCPacing is designed to adjust the dual variables in the dual percentile space, while constraining dual variables within the low-risk region through the pacing module using probabilistic throttling method.
\section{Risk-Constrained Pacing Algorithm}\label{section:rcpacing}
\begin{figure}[ht]
\centering
\includegraphics[width=0.45\textwidth]{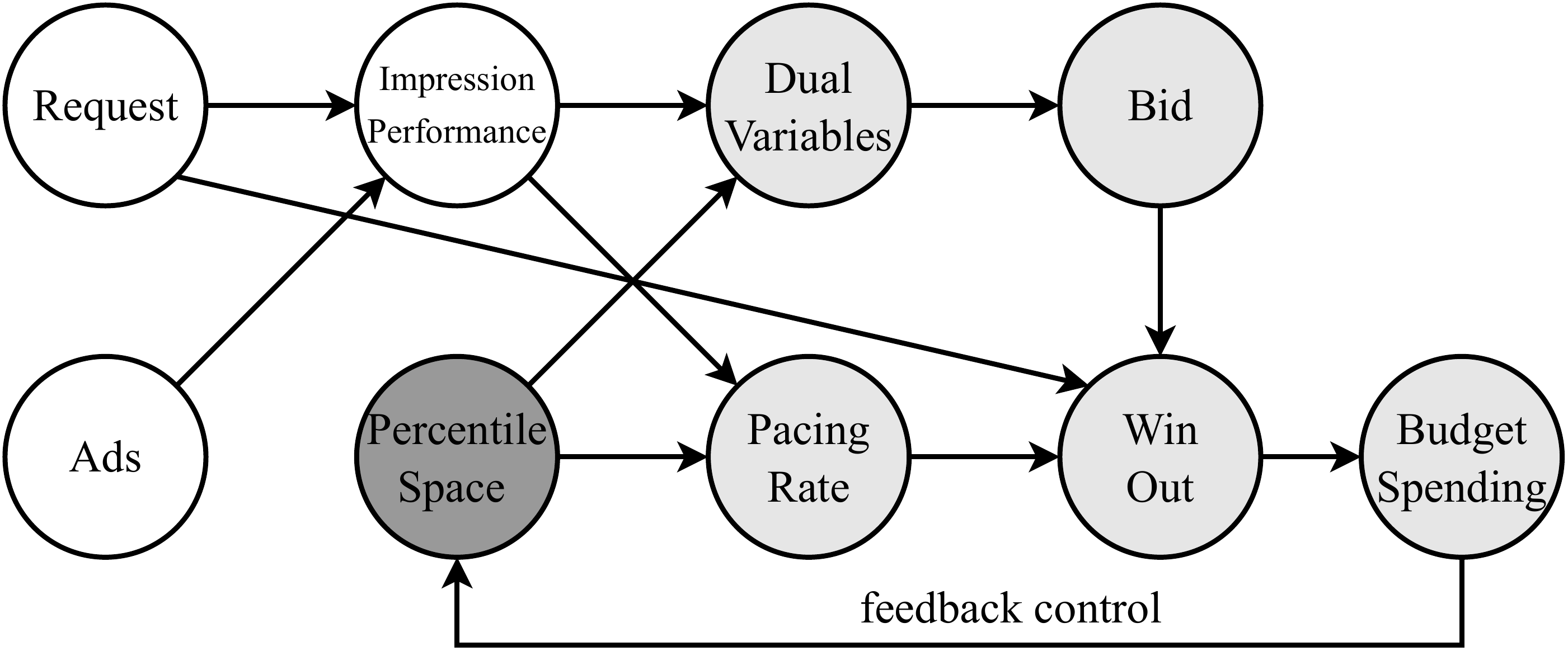}
\caption{Factor dependency graph of RCPacing.}
\label{fig:factor_dependency}
\end{figure}

The factor dependency of RCPacing is illustrated in figure \ref{fig:factor_dependency}. The dual variables and PTRs are adjusted in the dual percentile space of performance distributions. These two factors jointly determine the final win-out of each request. Although RCPacing adjusts the dual variables in percentile space rather than dual space, the Theorem 1 in appendix shows that it attains regret of order $O(\sqrt{T})$ when the length of the horizon $T$ and the initial number of resources are scaled proportionally.

\subsection{Parametric Percentile Transformation}
\subsubsection{Forward Transformation} RCPacing converts the CTR into the percentile space, which is called forward transformation, to assess the non-smooth risk and standardize the range of dual variables. Specifically, the CTR is first subjected to statistical Box-Cox transformation to achieve a normal shape, after which it is converted into the percentile space using the normal cumulative distribution function $\Phi(x)$. The parameter $\lambda_j^*$ of campaign $j$ can be estimated from global or campaign's historical logs using the maximum-likelihood method \cite{sakia1992box}:
\begin{equation}
\label{eq:boxcox_mle}
\lambda_j^*=\underset{\lambda_j}{\operatorname{argmax}} M L E\left(\lambda_j, v_{i j}\right)
\end{equation}

And Box-Cox transformation can be denoted as:
\begin{equation}
\label{eq:boxcox_trans}
v_{ij}^{boxc}=  BoxCox(\lambda_j^*,v_{ij})=
    \begin{cases}
        \frac{v^{\lambda_j^*}_{ij}-1}{\lambda_j^*} & \text { if } \lambda_j^*\neq 0 \\ \ln \left(v_{ij}\right) & \text { if } \lambda_j^*=0
    \end{cases}
\end{equation}

The mean $\mu_j$ and standard deviation $\sigma_j$ can be estimated:
\begin{equation}
\label{eq:boxcox_avg_std}
\mu_j=\mathbb{E}(v_{ij}^{boxc}),\ \sigma_j=\sqrt{\mathbb{E}\left[(v_{ij}^{boxc}-\mu_j)^2\right]}
\end{equation}

To improve the robustness of drifts in the user arrival distribution, RCPacing skews the transformation towards the middle percentile region by a factor $\epsilon$:
\begin{equation}
\label{eq:forward_trans}
\bar{v}_{ij} = \Phi\left( \frac{BoxCox(\lambda_j^*,v_{ij}) - \mu_j}{\sigma_j+\epsilon\sigma_j} \right) ,where\ \epsilon \ge 0
\end{equation}

\begin{figure}[t]
\centering
\includegraphics[width=0.48\textwidth]{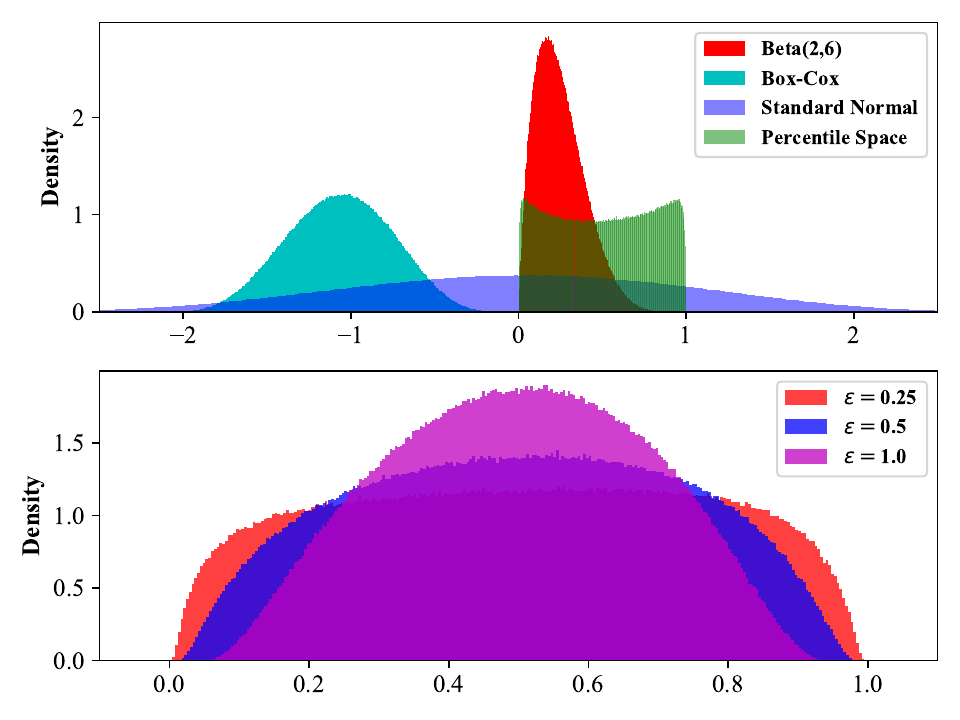}
\caption{The transformation process from beta distribution to percentile uniform distribution and the different skewness of the distribution under different $\epsilon$.}
\vspace{-2ex}
\label{fig:boxcox}
\end{figure}

\subsubsection{Backward Transformation} RCPacing periodically updates the dual variables in the percentile space through feedback and then performs a backward transformation of the percentile variables $\bar{\alpha}_j$ into the original dual space $\alpha_{j}$ for online service. It guarantees that RCPacing approaches the optimal solution in the original space rather than percentile space. Here is the backward process:
\begin{equation}
\label{eq:backward_trans}
\alpha_{j} = BoxCox^{-1}\left(\lambda_j^*,\mu_j + \Phi^{-1}(\bar{\alpha}_j)*(\sigma_j+\epsilon\sigma_j)\right)
\end{equation}

\subsection{Pacing Rate Factor Decoupling}
The pacing rate serves multiple functions in RCPacing, including constraining the percentile of dual variables within the safety region and addressing unexpected environmental changes and cold-start problem. RCPacing decouples the pacing rate into different factors to achieve optimal performance, for the campaign $j$ retrieved in request $i$:
\begin{equation}
\label{eq:ptr_perc}
\overline{PTR}_{ij} = PTR^{base}_{j} \cdot fp\left(\bar{\alpha}_j\right) \cdot fv\left(\bar{\alpha}_j,\bar{v}_{ij}\right)
\end{equation}
where $PTR^{base}_{j}$ is the basic statistical PTR, $f(\cdot)$ and $fv(\cdot)$ are the fine-tune factors. Given a  safe upper bound of percentile threshold $P_{ub}$ (such as 90\%), the expected PTR can be calculated based on its targeted audience $TA_j$ without considering the competition from other campaigns:
\begin{equation}
\label{eq:ptr_expect}
PTR^{exp}_j=\frac{B_j}{(1.0-P_{ub})TA_j}
\end{equation}

The initial value of $\bar\alpha_{j}$ can be expressed as:
\begin{equation}
\label{eq:ptr_alpha_init}
\bar\alpha^{(0)}_{j} =\left\{\begin{array}{l}
\begin{aligned}
P_{ub}&, \text{if } PTR^{exp}_j \le 1 \\
1 - (1-P_{ub})PTR^{exp}_j&, \text {otherwise.}
\end{aligned}
\end{array}\right.
\end{equation}

Given the global hyper-parameter $WR_{glb}$ (such as 0.2), the basic PTR considering the competition of $WR$ can be expressed as:
\begin{equation}
\label{eq:ptr_base_init}
PTR^{base}_j=\min \left\{1.0,PTR^{exp}_j/WR_{glb} \right\}
\end{equation}

\begin{figure}[t]
\centering
\includegraphics[width=0.48\textwidth]{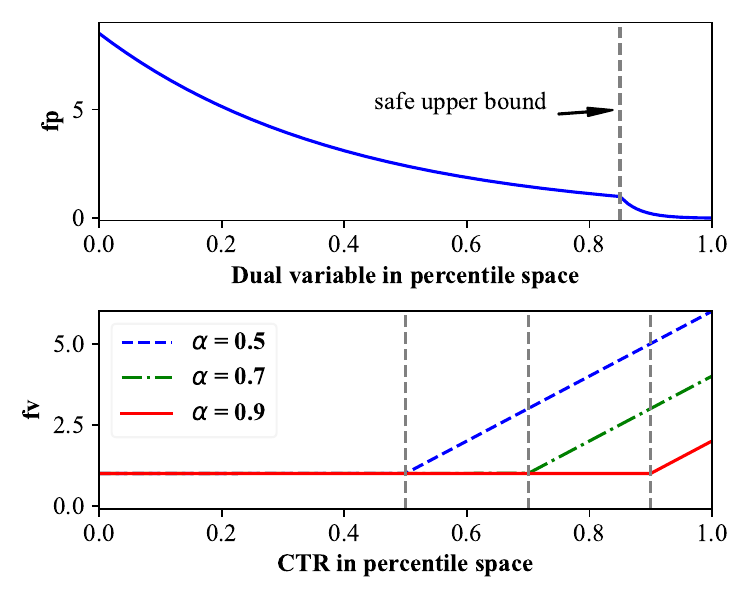}
\caption{Functions of $fp$ and $fv$ in percentile space.}
\vspace{-2ex}
\label{fig:fpfv}
\end{figure}

During the dynamic update in RCPacing, $PTR_j$ should be gradually increased to enhance traffic supply if $\bar{\alpha}_j<P_{ub}$. Conversely, it should be quickly decayed to reduce the non-smooth risk. It is illustrated in equation \ref{eq:ptr_fp} and figure \ref{fig:fpfv}:
\begin{equation}
\label{eq:ptr_fp}
fp\left(\bar{\alpha}_j\right) =\left\{\begin{array}{l}
\begin{aligned}
50^{(P_{ub}-\bar{\alpha}_j)/P_{ub}}&, \text{if } \bar{\alpha}_j \le P_{ub} \\
0.2^{(P_{ub}-\bar{\alpha}_j)/(P_{ub}-1)}&, \text { otherwise.}
\end{aligned}
\end{array}\right.
\end{equation}

Taking inspiration from smart pacing, RCPacing assigns a higher PTR to traffic with higher performance scores. Instead of employing discrete layered pacing, RCPacing utilizes linear functions to achieve non-uniform pacing:
\begin{equation}
\label{eq:ptr_fv}
fv\left(\bar{\alpha}_j,\bar{v}_{ij}\right)= 10(\bar{v}_{ij}-\bar{\alpha}_j)+1
\end{equation}

\subsection{Emergence Control and Cold Start Problem}
Despite RCPacing's adaptive adjustment of the PTR, it cannot completely mitigate the risks of non-smooth delivery caused by unpredictable factors, such as sharp increases in user traffic, significant distribution changes due to the switch of online real-time prediction models, offsets caused by updates of Box-Cox parameters, and modifications of budgets. Additionally, due to the absence of historical logs, there is also a risk of non-smoothness during the cold start phase. To address the these risks, RCPacing incorporates an emergent PTR intervention module (ePTR) that can be activated in emergency situations. The final PTR can be denoted as:
\begin{equation}
\label{eq:final_ptr}
PTR_{ij} = \min\{1,\overline{PTR}_{ij}\} \times ePTR_j
\end{equation}

The motivation behind ePTR is to limit the consumption speed within a certain range when a campaign is over-accelerated while maintaining the gradient direction of dual variables. The ratio of the actual cost to the expected cost can represent the spending speed of campaign $j$ during period $t$:
\begin{equation}
spd_j^{(t)} = \frac{Cost_j^{(t)}}{eCost_j^{(t)}}
\end{equation}

RCPacing uses proportional control instead of gradient methods to quickly control the risks. Given a safe upper ratio 2.0, the update of ePTR is:
\begin{equation}
\label{eq:udate_eptr}
ePTR_j^{(t+1)} =  \min\{1, ePTR_j^{(t)} * \min\{2, \frac{2}{spd_j^{(t)}}\} \}
\end{equation}

An initial trial rate is usually set for each campaign at the start of delivery to reduce the risks of the cold start problem.

\subsection{Adaptive Gradient Clipping}
Stable online iterative update of dual variables is also a critical factor for smooth delivery. However, choosing inappropriate learning rates can result in significant fluctuations and may have a cascading effect on the overall competitive environment. A simple and direct method is to restrict the change range into $\hat{\alpha}$ by gradient clipping \cite{chen2020gradient_clipping}. Given the updated dual variables $\tilde{\alpha}^{(t+1)}_{j}$, gradient clipping can be denoted as:
\begin{equation}
\bar{\alpha}^{(t+1)}_{j} = \max \left\{\bar{\alpha}^{(t)}_{j} - \hat{\alpha},\min\left\{\tilde{\alpha}^{(t+1)}_{j}, \bar{\alpha}^{(t)}_{j} + \hat{\alpha}\right\} \right\}
\end{equation}

Suppose that $spd_j^{(t)} < 1.0$ , which indicates that the campaign’s spending is lower than expected. The feedback control method will decrease the value of $\alpha_j^{(t)}$ to $\alpha_j^{(t+1)}$, leading to an increase in the bid price. Assuming that the competition remains the same, which indicates that $WR_{ij}^{(t+1)} \ge WR_{ij}^{(t)}, \text{ if } v_{ij}^{(t+1)}=v_{ij}^{(t)}$. Suppose the expected spending speed in the next period is equal to 1, it can be deduced that:
\begin{equation}
\begin{aligned}
&1.0= \frac{Cost_j^{(t+1)}}{eCost_j^{(t+1)}} =\frac{Cost_j^{(t+1)}}{eCost_j^{(t)}}=spd_j^{(t)}\frac{Cost_j^{(t+1)}}{Cost_j^{(t)}}\\
& = spd_j^{(t)}\frac{\mathbb{E}\left[RR_j^{(t+1)}\right] \sum PTR_j^{(t+1)} PR_j^{(t+1)} WR_j^{(t+1)}}{\mathbb{E}\left[RR_j^{(t)}\right] \sum PTR_j^{(t)} PR_j^{(t)} WR_j^{(t)}} \\
&\ge spd_j^{(t)} \sum PTR_j^{(t+1)} PR_j^{(t+1)} /\sum PTR_j^{(t)} PR_j^{(t)}\\
& = spd_j^{(t)} \mathbb{E}\left[PTR_j^{(t+1)} PR_j^{(t+1)}\right]/\mathbb{E}\left[PTR_j^{(t)} PR_j^{(t)}\right] \\
\end{aligned}
\end{equation}

Without consideration the effect of $ePTR$, $PTR$ and $PR$ are determined and have a monotonic decreasing relationship with $\bar{\alpha}$. We can calculate the expectation using the importance sampling method in uniform percentile space:
\begin{equation}
\begin{aligned}
\psi_j(\bar{\alpha}_j^{(t)}) &= \mathbb{E}\left[PTR_j^{(t)} PR_j^{(t)}\right] = \int_0^1 PTR_j^{(t)} (\bar{\alpha}_j^{(t)},\\&x ) \cdot PR_j(\bar{\alpha}_j^{(t)},x) \mathrm{d} x, \text{ } \ x \sim \text{uniform}(0,1)
\end{aligned}
\end{equation}

\begin{figure}[ht]
\centering
\includegraphics[width=0.48\textwidth]{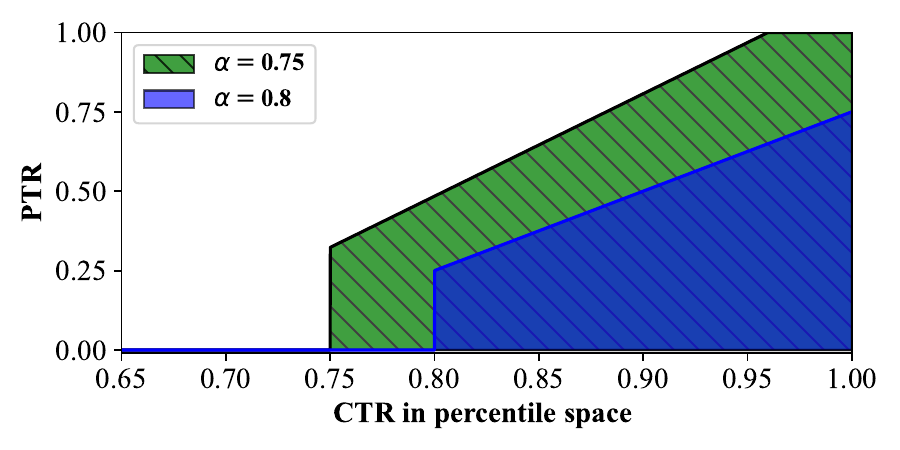}
\caption{The areas of the color section represent the value of $\psi_j(\bar{\alpha}_j^{(t)})$ under different variables $\alpha_j^{(t)}$.}
\vspace{-2ex}
\label{fig:grad_clip}
\end{figure}

The lower bound of $\alpha_j^{(t+1)}$ can be represented as:
\begin{equation}
\bar{\alpha}_j^{(t+1)} \ge \psi_j^{-1}\left(\psi_j(\bar{\alpha}_j^{(t)})/spd_j^{(t)}\right) = \psi_j^{-1}
\end{equation}
where $y=\psi_j^{-1}(\bar{\alpha}_j, x)$ can be approximated through an iterative process by solving the equation $\psi_j(\bar{\alpha}_j, y)=x$ based on the bisection method illustrated in figure \ref{fig:grad_clip}. To include $spd_j^{(t)} \ge 1.0$, $\bar{\alpha}_j^{(t+1)}$ should satisfy the following conditions:
\begin{equation}
\label{eq:dynamic_clip}
\bar{\alpha}^{(t+1)}_{j} =
\left\{\begin{array}{l}
\begin{aligned}
\max\left\{\tilde{\alpha}^{(t+1)}_{j},\bar{\alpha}^{(t)}_{j} - \hat{\alpha},\psi_j^{-1}\right\}&,\text { if } \tilde{g}_j^{(t)} \ge 0 \\
\min\left\{\tilde{\alpha}^{(t+1)}_{j},\bar{\alpha}^{(t)}_{j} + \hat{\alpha},\psi_j^{-1}\right\}&, \text { otherwise.}
\end{aligned}
\end{array}\right.
\end{equation}

\subsection{Bregman Divergence Selection}
Algorithm \ref{alg: DMD} presents the basic decision process based on Bregman divergence with respect to a given convex reference function. It is obvious that if we use the squared loss function and the dual update becomes:
\begin{equation}
h(\alpha)=\alpha^2 \Rightarrow \tilde{\alpha}^{(t+1)}_{j} = \bar{\alpha}_j^{(t)} - \eta \tilde{g}_j^{(t)},\forall j
\end{equation}

However, due to the higher fluctuation of PR in the high percentile region with the same shift, the variation magnitude of dual variables should be smaller to minimize non-smooth risk. It means that as $\bar{\alpha}_j$ approaches 1.0, the $\eta$ should become smaller. We propose a modified Itakura-Saito divergence \cite{banerjee2005clustering} to achieve this objective:
\begin{equation}
\label{eq:itakura}
\begin{aligned}
&h(\alpha)=-ln(1.5-\alpha) \Rightarrow \\
&\tilde{\alpha}^{(t+1)}_{j}=\bar{\alpha}^{(t)}_j-\frac{(1.5-\bar{\alpha}^{(t)}_j)^2}{1-\eta \tilde{g}_j^{(t)}(1.5-\bar{\alpha}^{(t)}_j)}\eta \tilde{g}_j^{(t)},\forall j \\
&\text{ where } \eta \tilde{g}_j^{(t)}(1.5-\bar{\alpha}^{(t)}_j) < 1
\end{aligned}
\end{equation}

The overall processing details of RCPacing are described in Algorithm \ref{alg: rcpacing}.

\begin{algorithm}[!t]
\caption{RCPacing}
\label{alg: rcpacing}
\begin{algorithmic}[1]
\REQUIRE Budget of the campaigns $\boldsymbol{B}$, safe upper bound $P_{ub}$, global win rate $WR_{glb}$, skew factor $\epsilon$, step size $\eta$, static gradient clipping $\hat{\alpha}$, total time period $T$

\STATE Budget exhausted campaign set $\mathcal{G} = \emptyset$
\STATE Calculate $\boldsymbol{PTR}^{base}$ and $\boldsymbol{\bar{\alpha}}^{(0)}$ with eq. \ref{eq:ptr_expect} $\sim$ \ref{eq:ptr_base_init}

\FOR{$t=0$ \textbf{to} $T-1$}

\STATE Estimate $\boldsymbol{\lambda}^*$, $\boldsymbol{\mu}$, and $\boldsymbol{\sigma}$ from historical logs with eq. \ref{eq:boxcox_mle} $\sim$ \ref{eq:boxcox_avg_std}

\STATE Obtain $\boldsymbol{\alpha}^{(t)}$ from $\boldsymbol{\bar{\alpha}}^{(t)}$ by backward transformation in eq. \ref{eq:backward_trans}

\STATE Receive $\boldsymbol{v}^{(t)}$ from online requests


\STATE Obtain $\boldsymbol{\bar{v}}^{(t)}$ from $\boldsymbol{v}^{(t)}$ by forward transformation in eq. \ref{eq:forward_trans}

\STATE Calculate $\boldsymbol{PTR}^{(t)}$ with eq. \ref{eq:ptr_perc} and eq. \ref{eq:ptr_fp} $\sim$ \ref{eq:final_ptr}

\STATE $\boldsymbol{bid}^{(t)} = \boldsymbol{v}^{(t)} - \boldsymbol{\alpha}^{(t)}$

\STATE Element-wise randomly set ${bid}^{(t)}_{ij} = 0$ with probability $1-{PTR}^{(t)}_{ij}$ and set ${bid}^{(t)}_{ij} = 0$ if ${j} \in \mathcal{G}$






\STATE $\boldsymbol{j}^* = \mathop{\arg\max}\left\{\boldsymbol{bid}^{(t)}\right\}$

\STATE Make the decision $\tilde{\boldsymbol{x}}^{(t)}$, where
\begin{equation}
    \tilde{{x}}^{(t)}_{ij} = \left\{
        \begin{aligned}
            1, & \; \text{if } {bid}^{(t)}_{ij} > 0 \text{ and } j=\boldsymbol{j}^*_i\\
            0,& \; \text{otherwise.} \\
        \end{aligned}
    \right.
\end{equation}

\STATE $\boldsymbol{B} = \boldsymbol{B} - \sum_i\tilde{\boldsymbol{x}}^{(t)}$

\STATE Add budget exhausted campaign to $\mathcal{G}$
 \STATE Calculate $\tilde{\boldsymbol{\alpha}}^{(t+1)}$ with eq. \ref{eq:itakura}

\STATE Update $\bar{\boldsymbol{\alpha}}^{(t+1)}$ by clipping $\tilde{\boldsymbol{\alpha}}^{(t+1)}$ with eq. \ref{eq:dynamic_clip}

\STATE Update $\boldsymbol{ePTR}^{(t+1)}$ with eq. \ref{eq:udate_eptr}
\ENDFOR
\end{algorithmic}
\end{algorithm}
\section{Experimental Results}\label{section:results}
This section begins with an introduction to the evaluation metrics and the baseline methods, and compares RCPacing to the baselines through offline and online experiments. 

\subsection{Evaluation Metrics}
\begin{itemize}
    \item \textbf{Delivery rate} is defined as the ratio of allocated impressions to the total budgets of the advertisers:
    \begin{equation}
        delivery \; rate = \frac{\sum_t\sum_{j}\tilde{x}_{j}^{(t)}}{\sum_jB_j}
    \end{equation}
    \item \textbf{Unsmoothness index} (UI) measures the deviation between the actual and expected budget consumption:
    \begin{equation}
        unsmoothness=\frac{1}{M}\sum_{j=1}^M\sqrt{\frac{1}{T}\sum_{t=0}^{T-1}\left(\tilde{x}_{j}^{(t)}-\rho_j\right)^2}
    \end{equation}
    \item \textbf{Average CTR} reflects the quality of impressions, is calculated as the ratio of clicks to the total impressions:
    \begin{equation}
        CTR_{avg} = \frac{\sum_t\sum_{j}v_{j}^{(t)}\tilde{x}_{j}^{(t)}}{\sum_t\sum_{j}\tilde{x}_{j}^{(t)}}
    \end{equation}
\end{itemize}

\subsection{Baseline Methods}
We compare RCPacing with the following four methods:
1) \textbf{DMD} \cite{balseiro2020dual} is a Lagrangian dual-based online allocation framework that maximizes revenue while adhering to resource constraints by adjusting their virtual bids. 
2) \textbf{Smart} Pacing \cite{xu2015smart} is a control-based method proposed to achieve smooth delivery and optimal performance by probabilistic throttling. 
3) \textbf{AUAF} \cite{cheng2022adaptive} is a dual-based method that optimizes delivery rate and impression quality with a fixed smoothness term. The dual variables are updated by feedback control algorithm to ensure fairness. 
4) \textbf{PDOA} \cite{zhou2021primal} solves online matching in dynamic environments with experts and meta-algorithm. It achieves smoothness by bid modification.
\subsection{Offline Evaluation}

\begin{table}[t]
\centering
\caption{Optimal values for the important hyper-parameters}
\label{table:params}
\renewcommand{\arraystretch}{1.25}
\begin{tabular}{c|cc}
\hline
parameter & value & description\\
\hline
$\epsilon$ & 0.1 & skew factor\\
$\eta$ & 0.2 & step size\\
$\hat{\alpha}$ & 0.05 & static gradient clipping  \\
$P_{ub}$ & 90\% & safe percentile upper bound\\
$WR_{glb}$ & 15\% & global win rate\\

\hline
\end{tabular}
\end{table}

\subsubsection{Datasets}
We construct a large-scale industrial dataset\footnote{Dataset and the code for all methods are available in https://github.com/danifree/RCPacing.} by collecting real-world ad-serving data from our display advertising system, which consists of 600K impressions and 300 GD ads. The impressions are evenly distributed across 50 time periods. The CTR values predicted by a DNN are reserved to measure the impression quality.

\subsubsection{Implementation Details}
Table \ref{table:params} provides a summary of the optimal values for the important hyper-parameters.

\subsubsection{Evaluation Results}
In order to exclude the influence of accidental factors, we randomly scale the budget of GD ads by a factor ranging from 0.8 to 1.2, and calculate the mean and standard deviation across 50 rounds. As shown in Table \ref{table:offline}, Smart Pacing achieves the highest average CTR, but its low delivery rate is inappropriate for GD allocation, which results in publishers being penalized for unsatisfied demand. RCPacing demonstrates a significant reduction in UI, with a 59.4\% and 50.8\% improvement compared to PDOA and AUAF, respectively. Furthermore, it delivers superior CTR performance, achieving a 23.1\% and 45.1\% increase compared to PDOA and AUAF.

\begin{table}[t]
\centering
\caption{Offline evaluation results}
\label{table:offline}
\renewcommand{\arraystretch}{1.25}
\resizebox{\columnwidth}{!}{
\begin{tabular}{c|ccc}
\hline
Method & Unsmoothness & Delivery Rate (\%) & CTR (\%)\\
\hline
DMD & $15.71\pm1.46$ & $\mathbf{100.0\pm0.0}$ & $5.39\pm0.02$\\
Smart & $10.52\pm1.04$ & $95.9\pm0.9$ & $\mathbf{7.88\pm0.37}$\\
AUAF & $12.95\pm1.29$ & $100.0\pm0.0$ & $5.14\pm0.01$\\
PDOA & $15.70\pm2.27$ & $100.0\pm0.0$ & $6.06\pm0.27$\\
RCPacing & $\mathbf{6.37\pm0.72}$ & $99.8\pm0.1$ & $7.46\pm0.44$\\
\hline
\end{tabular}
}
\end{table}

\begin{figure}[t]
\centering

  \vspace{-2ex}
\includegraphics[width=\columnwidth]{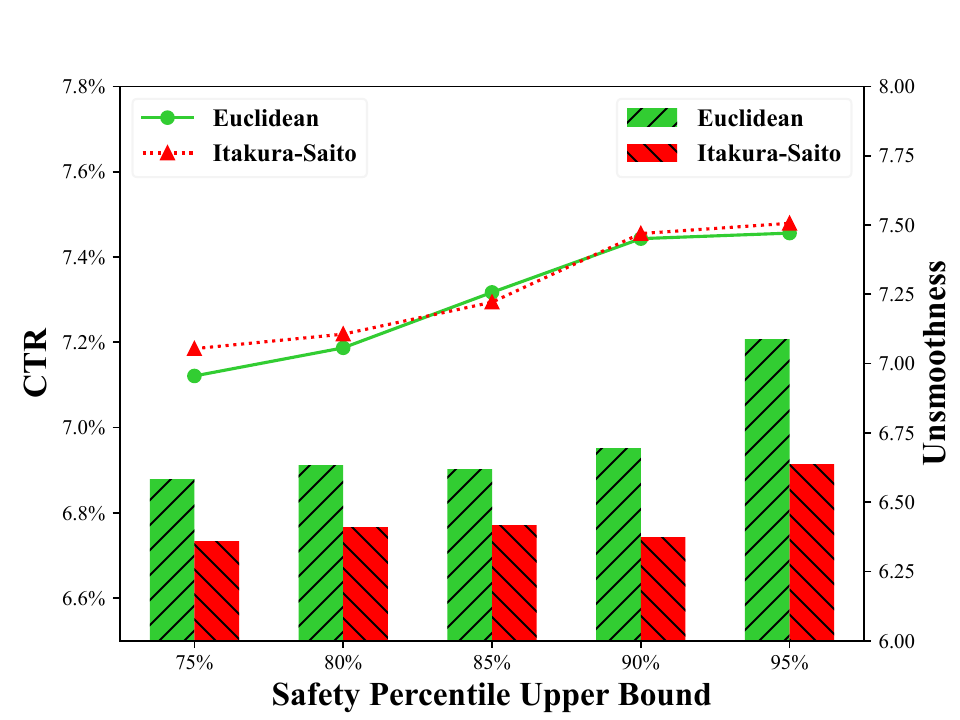}
\caption{The ablative analysis.}
\vspace{-2ex}
\label{fig:ablative}
\end{figure}

\subsubsection{Ablative Analysis}
We focus on UI and CTR since the delivery rates of the variants are very close to 100\%.
\begin{itemize}
    \item \textbf{The impact of percentile upper bound}: A higher safety percentile upper bound ($P_{ub}$) allows advertisers to filter low-quality impressions more effectively, but it also raises the risk of fluctuations. As demonstrated in figure \ref{fig:ablative}, RCPacing has higher CTR when using a larger $P_{ub}$, but there is a 4.14\% increase in unsmoothness when $P_{ub}$ is changed from 90\% to 95\% (Itakura-Saito divergence).
    \item \textbf{The impact of different divergence}: As mentioned earlier, a modified Itakura-Saito divergence helps alleviate the issue of high fluctuations in the high percentile range. Figure \ref{fig:ablative} illustrates that the proposed Itakura-Saito divergence provides better UI especial when $P_{ub}$ is high (e.g., a 3.74\% improvement in smoothness when $P_{ub}$ equals 90\%), while the average CTR is comparable to that of the Euclidean divergence.
\end{itemize}
Additional ablative analysis can be found in the appendix.

\subsection{Online Evaluation}

\begin{figure}[t]
\centering
\includegraphics[width=\columnwidth]{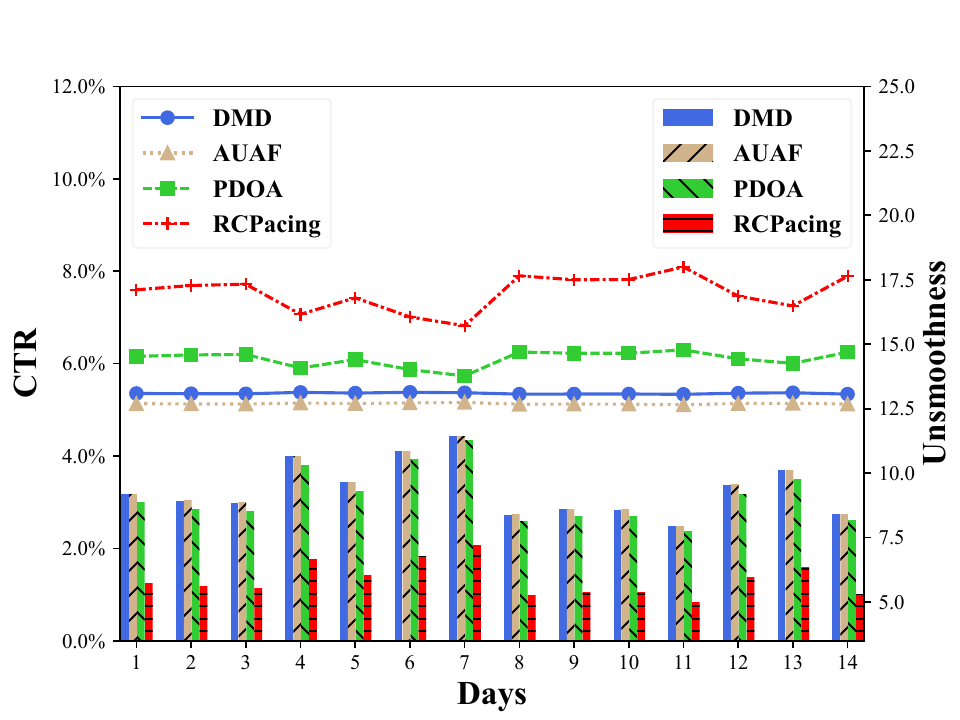}
\caption{The online evaluation results.}
\vspace{-2ex}
\label{fig:online}
\end{figure}

\subsubsection{Implementation Details}
In order to evaluate the performance of RCPaing in an online environment, we conduct A/B testing on our Taobao brand advertising platform for a continuous period of two weeks. Since the delivery rate for Smart pacing is too low for GD allocation, we only compare our method with DMD, AUAF, and PDOA. 

\subsubsection{Evaluation Results} As the delivery rates of all methods exceed 99.5\%, we concentrate on the other two metrics in figure \ref{fig:online}, RCPacing outperforms all the baselines. For example, compared with PDOA, our method achieves a 35.3\% and 23.4\% improvement in UI and CTR, respectively.
\section{Conclusion}\label{section:Conclusion}
GD contracts are a crucial source of revenue for large publishers. This paper presents a robust percentile risk-constrained pacing framework designed from the perspective of a publisher. RCPacing achieves smooth and optimal allocation for GD campaigns by leveraging its compatibility with the guaranteed allocation mechanism. Our analysis presents the relationship between non-smooth risks and percentile of dual variables, and RCPacing is designed to constrain dual variables within the low-risk region. Adaptive gradient clipping and modified Bregman divergence techniques are also employed to achieve more stable update of dual variables. We also illustrate the trade-off and flexible control over smooth and optimal allocation in online matching. Our experimental evaluations on real-world A/B testing demonstrate that RCPacing outperforms other compared methods and it has been widely deployed in Taobao display advertising system.

\bibliography{main}

\clearpage
\section{Appendix}

\begin{table}[h]
\centering
\caption{The impact of slope $k$}
\label{table:fv}
\renewcommand{\arraystretch}{1.25}
\resizebox{\columnwidth}{!}{
\begin{tabular}{c|ccc}
\hline
$k$ & Unsmoothness & Delivery Rate (\%) & CTR (\%)\\
\hline
0 & $6.54\pm0.65$ & $99.8\pm0.1$ & $7.17\pm0.39$\\
1 & $6.68\pm0.65$ & $99.8\pm0.1$ & $7.36\pm0.37$\\
10 & $6.37\pm0.72$ & $99.8\pm0.1$ & $7.46\pm0.44$\\
100 & $6.84\pm0.61$ & $99.9\pm0.0$ & $7.71\pm0.35$\\
1000 & $7.07\pm0.72$ & $99.9\pm0.1$ & $7.77\pm0.42$\\
\hline
\end{tabular}
}
\end{table}
\begin{table}[h]
\centering
\caption{The impact of gradient clipping}
\label{table:grad_clip}
\renewcommand{\arraystretch}{1.25}
\resizebox{\columnwidth}{!}{
\begin{tabular}{c|c|ccc}
\hline
Method & $\eta$ & Unsmoothness & Delivery Rate (\%) & CTR (\%)\\
\hline
\multirow{3}{*}{w/o clip} & 0.2 & $6.57\pm0.59$ & $99.8\pm0.1$ & $7.44\pm0.39$\\
& 0.4 & $7.68\pm0.82$ & $99.7\pm0.1$ & $7.30\pm0.39$\\
& 0.8 & $11.54\pm1.13$ & $99.7\pm0.0$ & $7.41\pm0.41$\\
\hline
\multirow{3}{*}{w/ clip} & 0.2 & $6.37\pm0.72$ & $99.8\pm0.1$ & $7.46\pm0.44$\\
& 0.4 & $7.16\pm0.79$ & $99.8\pm0.1$ & $7.32\pm0.43$\\
& 0.8 & $8.26\pm0.72$ & $99.9\pm0.1$ & $7.36\pm0.39$\\
\hline
\end{tabular}
}
\end{table}

\subsection{Supplementary experiments}
This subsection presents additional experimental results that further demonstrate the efficacy of RCPacing's design strategies.
\subsubsection{The impact of performance-based pacing}
The performance-based pacing ($fv=k(\bar{v}_{ij}-\bar{\alpha}_j)+1$) assigns a higher PTR to traffic with higher performance scores, and the slope $k$ determines how significant this non-uniform is. As shown in Table \ref{table:fv}, deactivating the performance-based pacing by setting $k=0$ yields the lowest average CTR. The average CTR increases with the increase in $k$, but it will eventually reach saturation since the PTR cannot exceed 1. 


\subsubsection{The impact of gradient clipping}
During iterative updates of the dual variables, gradient clipping is an effective technique that restricts the range of changes to prevent significant fluctuations caused by inappropriate learning rates. Table \ref{table:grad_clip} demonstrates that when the learning rate is excessively large, gradient clipping aims in maintaining a smooth and stable allocation.

\subsection{Theoretical proof}

\subsubsection{Assumption 1}
(Assumptions on constraint set $\mathcal{X}$). We assume that: (i) $0 \in \mathcal{X}$, and (ii) for each $x_t \in \mathcal{X} \text{ and } x_t \neq 0$, $x_t$ is a one-hot vector indicating which campaign the impression is assigned to.

The above assumption implies that we can only assign an impression to one campaign at a time. Moreover, we can always take the void action by choosing $x_t=0$ in order to make sure we do not exceed the budget constraints. This guarantees the existence of a feasible solution.

\subsubsection{Assumption 2}
(Assumptions on revenue function $f$).
\begin{itemize}
    \item $f\left(x\right)=v^{\top}x=\sum_{j=1}^Mv_{j}x_{j}$, $v_{j}$ is the impression quality and follows a beta distribution:
    \begin{equation}
        v_{j} \sim Beta(m,n), \quad m \ge 2, n \ge 2 
    \end{equation}
    \item There exists a positive constant $\bar{f}$ such that $f\left(x\right) \le \bar{f}$ is satisfied in all cases.
\end{itemize}

\subsubsection{Assumption 3}
(Assumptions on dual variable $\alpha$). We assume that $\alpha_{j} \in [0,1]$ is satisfied for any campaign $j$. Since campaign $j$ will not participate in the auction when $v_{j} \le \alpha_{j}$, it doesn't make sense for $\alpha_{j}$ to be greater than 1. Moreover, from the definition of the dual problem (equation \ref{eq:dual_problem}), we have $\alpha_{j} \ge 0$.  

\subsubsection{Assumption 4}
(Assumptions on budget parameter $\rho$). We assume there exist positive constants $\bar{\rho}$ and $\underline{\rho}$ such that $\underline{\rho} \le \rho_j \le \bar{\rho}$ is satisfied for any campaign $j$.

\subsubsection{Assumption 5}
(Assumptions on reference function $h\left(\cdot\right)$). We assume
\begin{itemize}
    \item $h\left(\bar{\alpha}\right)$ is coordinate-wisely separable, i.e., $h\left(\bar{\alpha}\right)=\sum_jh_j\left(\bar{\alpha}_j\right)$ where $h_j\left(\cdot\right)$ is a convex univariate function.
    \item $h\left(\bar{\alpha}\right)$ is $\sigma_1$-strongly convex in $l_1$-norm in $[0,1]$, i.e., $h\left(\bar{\alpha}_1\right) \ge h\left(\bar{\alpha}_2\right) + \left<\nabla h\left(\bar{\alpha}_2\right), \bar{\alpha}_1-\bar{\alpha}_2\right>+\frac{\sigma_1}{2}||\bar{\alpha}_1-\bar{\alpha}_2||^2_1$ for any $\bar{\alpha}_1,\bar{\alpha}_2 \in [0,1]$
    \item $h\left(\bar{\alpha}\right)$ is $\sigma_2$-strongly convex in $l_2$-norm in $[0,1]$, i.e., $h\left(\bar{\alpha}_1\right) \ge h\left(\bar{\alpha}_2\right) + \left<\nabla h\left(\bar{\alpha}_2\right), \bar{\alpha}_1-\bar{\alpha}_2\right>+\frac{\sigma_2}{2}||\bar{\alpha}_1-\bar{\alpha}_2||^2_2$ for any $\bar{\alpha}_1,\bar{\alpha}_2 \in [0,1]$
\end{itemize}

\subsubsection{Definition 1} 
\label{theory: def1}
We define $\alpha^{max} \in \mathbb{R}^M$ such that $\alpha^{max}_j := \frac{\bar{f}}{\rho_j} + 1$, where $M$ is the total number of campaigns.

\subsubsection{Definition 2}
We define the stopping time $\tau_j$ of campaign $j$ as the first time less than $T$ that satisfies
\begin{equation}
    \sum_{t=1}^{\tau_j}x_{tj}+1 \ge \rho_jT
\end{equation}
$\tau_j$ is a random variable, and campaign $j$ will adhere to its budget constraint until the stopping time $\tau_j$. To prevent campaign $j$ from participating in any auctions after its budget is exhausted, we set $v_{tj}$ to be $0$ after $\tau_j$.

Moreover, we define the stopping time $\tau_A$ of the algorithm as
\begin{equation}
    \tau_A = \max_j\{\tau_j\}
\end{equation}

\subsubsection{Definition 3}
We define the expected revenue of an algorithm $A$ as
\begin{equation}
    R(A) = \mathbb{E}\left[\sum_{t=1}^Tf_t\left(x_t\right)\right]
\end{equation}
where $x_t$ is the allocation decision made by the algorithm at time $t$. We take the offline problem as the baseline for comparison, which determines the optimal allocation under complete information of all requests and then calculates the expected revenue across all possible outcomes:
\begin{equation}
    \operatorname{OPT} = \mathbb{E}\left[
    \begin{split}
        \max_{x_t\in\mathcal{X}}\sum_{t=1}^Tf_t(x_t)=\sum_{t=1}^Tv_t^{\top}x_t \\
        \text{s.t.} \sum_{t=1}^Tx_t\le B = T \rho
    \end{split}\right]
\end{equation}

Moreover, the regret of algorithm $A$ is defined as:
\begin{equation}
    \operatorname{Regret}\left(A\right):=\operatorname{OPT}-R\left(A\right)
\end{equation}

\subsubsection{Lemma 1} 
\label{theory: lemma1}
Suppose $\alpha_j \in [0, 1], \forall j$, then there exists a constant $K_1$ such that $\bar{\alpha}_j \le K_1\alpha_j, \forall j$ always holds.

\noindent \textbf{Proof.} Denote the ratio $\bar{\alpha}_j / \alpha_j$ as
\begin{equation}
    \mathcal{R}_1\left(\alpha_j\right) = \frac{\int^{\alpha_j}_0\frac{1}{B\left(m,n\right)}s^{m-1}\left(1-s\right)^{n-1}ds}{\alpha_j}
\end{equation}
its derivative is obtained by
\begin{equation}
    \mathcal{R}'_1\left(\alpha_j\right)=\frac{\alpha^m_j\left(1-\alpha_j\right)^{n-1}-\int^{\alpha_j}_0s^{m-1}\left(1-s\right)^{n-1}ds}{\alpha^2_jB\left(m,n\right)}
\end{equation}

Let 
\begin{equation}
    \mathcal{G}_1\left(\alpha_j\right)=\alpha^m_j\left(1-\alpha_j\right)^{n-1}-\int^{\alpha_j}_0s^{m-1}\left(1-s\right)^{n-1}ds
\end{equation}
its derivative is obtained by
\begin{equation}
\mathcal{G}'_1\left(\alpha_j\right) = \alpha^{m-1}_j\left(1-\alpha_j\right)^{n-2}\left[m-1-\left(m+n-2\right)\alpha_j\right]
\end{equation}

Let $\alpha'_j = \left(m-1\right)/\left(m+n-2\right)$, $\mathcal{G}_1\left(\alpha_j\right)$ monotonically increases on $\left[0,\alpha'_j\right]$, and monotonically decreases on $\left[\alpha'_j,1\right]$. Since $\mathcal{G}_1\left(0\right)=0$, $\mathcal{G}_1\left(\alpha'_j\right)>0$, $\mathcal{G}_1\left(1\right)=-\int^1_0s^{m-1}\left(1-s\right)^{n-1}$,
there must be an $\alpha''_j \in \left[\alpha'_j,1\right]$ to make $\mathcal{G}_1\left(\alpha''_j\right)=0$. Thus $\mathcal{R}^{max}_1=\mathcal{R}_1\left(\alpha''_j\right)$. Let $K_1=\mathcal{R}^{max}_1$, we have $\bar{\alpha}_j =\mathcal{R}_1\left(\alpha_j\right)\alpha_j \le \mathcal{R}^{max}_1\alpha_j = K_1\alpha_j$, and the lemma is proved.

\subsubsection{Lemma 2} 
Suppose the update of dual variables in the original space is satisfied that $0 < \Delta_m \le \Delta \le 1$, then the percentile update must be satisfied that $\bar{\Delta} \ge K_2\Delta$, where $K_2$ is a positive constant.

\noindent \textbf{Proof.} Denote the ratio $\bar{\Delta} / \Delta$ as
\begin{equation}
    \mathcal{R}_2\left(\alpha_j,\Delta\right)=\frac{\int^{\alpha_j+\Delta}_{\alpha_j}\frac{1}{B\left(m,n\right)}s^{m-1}\left(1-s\right)^{n-1}ds}{\Delta} > 0
\end{equation}

Since $\mathcal{R}_2\left(\alpha_j,\Delta\right)$ is continuous when $\Delta \in [\Delta_m, 1]$ and $\alpha_j \in \left[0,1\right]$, there must be a positive constant $K_2$ satisfying $K_2 \le \min_{\alpha_j, \Delta}\left\{\mathcal{R}_2\left(\alpha_j,\Delta\right)\right\}$. Thus we have $\bar{\Delta} = \mathcal{R}_2\left(\alpha_j,\Delta\right)\Delta \ge K_2 \Delta$, and the lemma is proved.

\subsubsection{Lemma 3} 

Let $\tilde{g}=\nabla f^*\left(\alpha\right)+\rho$ with $f\left(x\right)=v^{\top}x$ and $v \in \left\{v_1, \cdots, v_n\right\}$, and $\bar{\alpha}^+=\mathop{\arg\min}_{\bar{\alpha}^* \ge 0}\left<\tilde{g},\bar{\alpha}^*\right>+\frac{1}{\eta}V_h\left(\bar{\alpha}^*,\bar{\alpha}\right)$. Suppose $\alpha \le \alpha^{max}$and $\eta \le K_1\sigma_2$, then it holds that $\bar{\alpha}^+ \le \bar{\alpha}^{max} = K_1\alpha^{max}$.

\noindent \textbf{Proof.} Denote $J := \left\{j|\bar{\alpha}^+_j > 0\right\}$, then we just need to show that $\bar{\alpha}^+_j \le K_1\alpha^{max}$ holds for any $j \in J$. Since we update the dual variables in the percentile space as
\begin{equation}
    \label{eq:update1}
    \bar{\alpha}^+ = \mathop{\arg\min}_{\bar{\alpha}^* \ge 0}\left<\tilde{g},\bar{\alpha}^*\right>+\frac{1}{\eta}V_h\left(\bar{\alpha}^*,\bar{\alpha}\right)
\end{equation}

\begin{equation}
    \label{eq:update2}
    V_h\left(\bar{\alpha}^*,\bar{\alpha}\right) = h\left(\bar{\alpha}^*\right)-h\left(\bar{\alpha}\right)-\left<\nabla h\left(\bar{\alpha}\right),\bar{\alpha}^*-\bar{\alpha}\right>
\end{equation}
it holds for any $j \in J$ that
\begin{equation}
    \label{eq:util_eq14}
    \dot{h}_j\left(\bar{\alpha}^+_j\right) = \dot{h}_j\left(\bar{\alpha}_j\right)-\eta\tilde{g}_j=\dot{h}_j\left(\bar{\alpha}_j\right)-\eta \left( \nabla f^*\left(\alpha\right)\right)_j-\eta\rho_j
\end{equation}

Define $h^*_j\left(c\right) = \max_{\bar{\alpha}}\left\{c\bar{\alpha}_j-h_j\left(\bar{\alpha}_j\right)\right\}$ as the conjugate function of $h_j\left(\bar{\alpha}_j\right)$, then by the property of conjugate function it holds that $h_j^*\left(\cdot\right)$ is a $\frac{1}{\sigma_2}$-smooth univariate convex function. Furthermore, $\dot{h}^*_j\left(\cdot\right)$ is increasing, and $\dot{h}^*_j\left(\dot{h}_j\left(\bar{\alpha}_j\right)\right) = \bar{\alpha}_j$.

Now define $\tilde{x}:=\mathop{\arg\max}_{x \in \mathcal{X}}\left\{f\left(x\right)-\alpha^{\top}x\right\}=-\nabla f^*\left(\alpha\right)$. Then it holds that $0=f\left(0\right) \le f\left(\tilde{x}\right)-\alpha^{\top}\tilde{x} \le \bar{f}-\alpha^{\top}\tilde{x}$, whereby $\alpha^{\top}\tilde{x} \le \bar{f}$. Since $\alpha \ge 0, \tilde{x}_j \in \left\{0,1\right\}$, it holds for any $j \in J$ that $\tilde{x}_j \le \min \left(\frac{\bar{f}}{\alpha_j}, 1\right)$. Together with equation \ref{eq:util_eq14}, it holds that
\begin{equation}
    \label{eq:util_eq15}
    \dot{h}_j\left(\bar{\alpha}^+_j\right) \le \dot{h}_j\left(\bar{\alpha}_j\right)+\eta \min \left(\frac{\bar{f}}{\alpha_j}, 1\right)-\eta\rho_j
\end{equation}

If $\frac{\bar{f}}{\rho_j} \le \alpha_j \le \alpha_j^{max}$, we have $\min\left(\frac{\bar{f}}{\alpha_j},1\right)-\rho_j \le 0$, thus it holds that $\bar{\alpha}_j^+ \le \bar{\alpha}_j \le K_1 \alpha_j \le K_1\alpha_j^{max}$ by utilizing equation \ref{eq:util_eq15}, Lemma 1, and convexity of $\dot{h}_j$. Otherwise, $\alpha_j \le \frac{\bar{f}}{\rho_j}$, and furthermore,
\begin{equation}
    \begin{aligned}
        \bar{\alpha}^+_j &= \dot{h}^*_j\left(\dot{h}_j\left(\bar{\alpha}^+_j\right)\right) \le \dot{h}^*_j\left(\dot{h}_j\left(\bar{\alpha}_j\right)+\eta\right) \\
        & \le \dot{h}^*_j\left(\dot{h}_j\left(\bar{\alpha}_j\right)\right)+\frac{\eta}{\sigma_2} \le K_1\alpha_j+\frac{\eta}{\sigma_2} \\
        & \le K_1 \left(\frac{\bar{f}}{\rho_j}+1\right)=K_1\alpha_j^{max}
    \end{aligned}
\end{equation}

where the first inequality is from equation \ref{eq:util_eq15} and the monotonicity of  $\dot{h}^*_j\left(\cdot\right)$, the second inequality is from $\dot{h}^*_j\left(\dot{h}_j\left(\bar{\alpha}_j\right)\right) = \bar{\alpha}_j$ and the $\frac{1}{\sigma_2}$-smoothness of $h^*_j\left(\cdot\right)$, the third inequality is from Lemma 1, and the last equality follows from Definition 1. This finishes the proof of the lemma.

\subsubsection{Proposition 1}

It holds for any $\alpha \geq 0$ that
\begin{equation}
\label{eq:ub_of_opt}
    \operatorname{OPT} \leq TD\left(\alpha\right).
\end{equation}

\noindent \textbf{Proof.} Notice that for any $\alpha \geq 0$, it holds that
\begin{equation}
\begin{aligned}
& \operatorname{OPT}\\
& =\mathbb{E}\left[\begin{array}{cl}
\max _{x_t \in \mathcal{X}} & \sum_{t=1}^T f_t\left(x_t\right) \\
\text { s.t. } & \sum_{t=1}^T x_t \leq T \rho
\end{array}\right] \\
& \leq \mathbb{E}\left[\max _{x_t \in X} \sum_{t=1}^T f_t\left(x_t\right)+T \alpha^{\top} \rho-\alpha^{\top} \sum_{t=1}^T x_t\right] \\
& =T \mathbb{E}\left[\max _{x \in X} f(x)-\alpha^{\top} x+\alpha^{\top} \rho\right] \\
& =T\left(\sum_{i=1}^n p_i \max _{x \in \mathcal{X}}\left\{f_i(x)-\alpha^{\top} x\right\}+\alpha^{\top} \rho\right) \\
& =T\left(\sum_{i=1}^n p_i f_i^*\left(\alpha\right)+\alpha^{\top} \rho\right) \\
&
\end{aligned}
\end{equation}
where the first inequality is because of the feasibility of $x$ and $\alpha \geq 0$ and the last equality is due to the definition of $f_i^*$. This finishes the proof.

\subsubsection{Proposition 2}

Consider Algorithm \ref{alg: rcpacing} with step-size $\eta \le K_1\sigma_2$. Then it holds that $\bar{\alpha}_t \le \bar{\alpha}^{max} = K_1\alpha^{max}$ for any $t \le T$. Furthermore, it holds with probability 1 that
\begin{equation}
    T-\tau_A \le \frac{1}{\eta\underline{\rho}}||\nabla h\left(K_1\alpha^{max}\right)-\nabla h\left(\bar{\alpha}_{0}\right)||_\infty + \frac{1}{\underline{\rho}}.
\end{equation}

\noindent \textbf{Proof.} First, a direct application of Lemma 3 shows that for any $t \le T$, $\bar{\alpha}_t \le \bar{\alpha}^{max} = K_1\alpha^{max}$. Next, it follows by the definition of $\tau_A$ that $\sum_{t=1}^{\tau_A}x_{tj}+1 \ge \rho_jT$ is satisfied for all $j$. By the definition of $\tilde{g}_t$, we have
\begin{equation}
    \sum_{t=1}^{\tau_A}\tilde{g}_{tj}=\rho_j\tau_A-\sum_{t=1}^{\tau_A}x_{tj} \le \rho_j\tau_A - \rho_jT + 1
\end{equation}
thus
\begin{equation}
\label{eq:util_eq16}
    T-\tau_A \le \frac{1-\sum_{t=1}^{\tau_A}\tilde{g}_{tj}}{\rho_j}, \forall j
\end{equation}
On the other hand, it follows the update rule (equation \ref{eq:update1} and \ref{eq:update2}) that for any $t \le \tau_A$,
\begin{equation}
    \dot{h}_j\left(\bar{\alpha}_{\left(t+1\right)j}\right) \ge \dot{h}_j\left(\bar{\alpha}_{tj}\right)-\eta\tilde{g}_{tj}
\end{equation}

Thus,
\begin{equation}
\label{eq:util_eq17}
    \begin{aligned}
\sum_{t=1}^{\tau_A}-\tilde{g}_{tj} &\le \frac{1}{\eta}\left[\dot{h}_j\left(\bar{\alpha}_{\left(\tau_A+1\right)j}\right)-\dot{h}_j\left(\bar{\alpha}_{0j}\right)\right]  \\
& \le \frac{1}{\eta}\left[\dot{h}_j\left(K_1\alpha_j^{max}\right)-\dot{h}_j\left(\bar{\alpha}_{0j}\right)\right]
\end{aligned}
\end{equation}
where the last inequality is due to the monotocity of $\dot{h}^*_j\left(\cdot\right)$. Combining equation \ref{eq:util_eq16} and \ref{eq:util_eq17}, we reach
\begin{equation}
    T-\tau_A \le \frac{\dot{h}_j\left(K_1\alpha_j^{max}\right)-\dot{h}_j\left(\bar{\alpha}_{0j}\right)}{\eta \rho_j} + \frac{1}{\rho_j}, \forall j
\end{equation}

This finishes the proof by noticing that $\rho_j \ge \underline{\rho}$ and $\dot{h}_j\left(K_1\alpha_j^{max}\right)-\dot{h}_j\left(\bar{\alpha}_{0j}\right) \le ||\nabla h\left(K_1\alpha^{max}\right)-\nabla h\left(\bar{\alpha}_{0}\right)||_\infty$.

\subsubsection{Proposition 3}
Consider the Algorithm \ref{alg: rcpacing} with given step size $\eta$ under Assumptions 1-5. Let $\tau_A$ be the stopping time defined in Definition 2. Denote is $\hat{\alpha}_{\tau_A}=\frac{\sum_{t=1}^{\tau_A}\alpha_t}{\tau_A}$. Then the following inequality holds:
\begin{equation}
\begin{aligned}
& \mathbb{E}\left[\tau_A D\left(\hat{\alpha}_{\tau_A}\right)-\sum_{t=1}^{\tau_A} f_t\left(x_t\right)\right] \\
& \leq \frac{2\left(1+\bar{\rho}^2\right)}{K_2\sigma_1} \eta \mathbb{E}\left[\tau_A\right]+\frac{V_h\left(0, \bar{\alpha}_0\right)}{K_2\eta}
\end{aligned}
\end{equation}

\noindent \textbf{Proof.} Before proving Proposition 3, we first introduce some new notations which are used in the proof. By the definition of conjugate function, we can rewrite the dual problem (equation \ref{eq:dual_problem}) as the following saddle-point problem:
\begin{equation}
\label{eq:saddle}
    (S): \min _{0 \leq \alpha} \max _{y \in p\mathcal{X}} L(y, \alpha):=\sum_{i=1}^n p_i f_i\left(y_i / p_i\right)-\alpha^{\top}By+\alpha^{\top} \rho
\end{equation}
where $y := \left[y_1,\ldots,y_n\right] \in \mathbb{R}^{nM}$, $B := \left[I_1;\ldots;I_n\right] \in \mathbb{R}^{M \times nM}$, $p\mathcal{X} := \left\{y \mid y_i \in p_i X\right\} \subseteq \mathbb{R}^{nM}_+$, and $I_i \in \mathbb{R}^{M \times M}$ is an identity matrix. By minimizing over $\alpha$ in equation \ref{eq:saddle}, we obtain the following primal problem:
\begin{equation}
    \begin{gathered}
        (P): \max _y P(y):=\sum_{i=1}^n p_i f_i\left(y_i / p_i\right) \\
        \text { s.t. } B y \leq \rho \\
        y \in p\mathcal{X}
    \end{gathered}
\end{equation}

The decision variable $y_i/p_i \in \mathcal{X}$ can be interpreted as the expected action to be taken when a request of type $i$ arrives. Therefore, (P) can be interpreted as a deterministic optimization problem in which resource constraints can be satisfied in expectation. Moreover, we define an auxiliary primal vari- able sequence $\left\{z_t\right\}_{{t=1}, \ldots, T}$:
\begin{equation}
\label{eq:util_eq22}
    z_t=\arg \max _{z \in p\mathcal{X}} L\left(z, \alpha_t\right)
\end{equation}

As a direct consequence of equation \ref{eq:saddle} and \ref{eq:util_eq22}, we obtain:
\begin{equation}
\label{eq:util_eq23}
    g_t:=-B z_t+\rho=\nabla_\alpha L\left(z_t, \alpha_t\right) \in \partial_\alpha D\left(\alpha_t\right)
\end{equation}

From the definition of $\tilde{g}_t$ and $\bar{\rho}$, we have
\begin{equation}
\label{eq:util_eq24}
    \mathbb{E}_{\gamma_t}\left\|\tilde{g}_t\right\|_{\infty}^2 \leq 2\left(\mathbb{E}_{\gamma_t}\left\|x_t\right\|_{\infty}^2+\|\rho\|_{\infty}^2\right) \leq 2\left(1+\bar{\rho}^2\right)
\end{equation}

Note that $\bar{\alpha}_t \in \sigma\left(\xi_{t-1}\right)$, $g_t \in \sigma\left(\xi_{t-1}\right)$, and $\tilde{g}_t \in \sigma\left(\xi_t\right)$, where $\sigma(X)$ denotes the sigma algebra generated by a stochastic process $X$. Notice $\mathbb{E}_{\gamma_t} \tilde{g}_t=g_t$, thus it holds for any $\bar{\alpha} \in [0, 1]$ that
\begin{equation}
    \begin{aligned}
& \left\langle g_t, \bar{\alpha}_t-\bar{\alpha}\right\rangle \\
= & \left\langle\mathbb{E}_{\gamma_t}\left[\tilde{g}_t \mid \bar{\alpha}_t\right], \bar{\alpha}_t-\bar{\alpha}\right\rangle \\
\leq &\mathbb{E}_{\gamma_t}  {\left[\left\langle\tilde{g}_t, \bar{\alpha}_t-\bar{\alpha}_{t+1}\right\rangle+\frac{1}{\eta} V_h\left(\bar{\alpha}, \bar{\alpha}_t\right)\right.} \\
& \left.\quad-\frac{1}{\eta} V_h\left(\bar{\alpha}, \bar{\alpha}_{t+1}\right)-\frac{1}{\eta} V_h\left(\bar{\alpha}_{t+1}, \bar{\alpha}_t\right) \mid \bar{\alpha}_t\right] \\
\leq & \mathbb{E}_{\gamma_t}\left[\left\langle\tilde{g}_t, \bar{\alpha}_t-\bar{\alpha}_{t+1}\right\rangle+\frac{1}{\eta} V_h\left(\bar{\alpha}, \bar{\alpha}_t\right)\right. \\
& \left.\quad-\frac{1}{\eta} V_h\left(\bar{\alpha}, \bar{\alpha}_{t+1}\right)-\frac{\sigma_1}{2 \eta}\left\|\bar{\alpha}_{t+1}-\bar{\alpha}_t\right\|_1^2 \mid \bar{\alpha}_t\right] \\
\leq & \mathbb{E}_{\gamma_t}\left[\frac{\eta}{\sigma_1}\left\|\tilde{g}_t\right\|_{\infty}^2+\frac{1}{\eta} V_h\left(\bar{\alpha}, \bar{\alpha}_t\right)-\frac{1}{\eta} V_h\left(\bar{\alpha}, \bar{\alpha}_{t+1}\right) \mid \bar{\alpha}_t\right] \\
\leq & \frac{2 \eta}{\sigma_1}\left(1+\bar{\rho}^2\right)+\frac{1}{\eta} V_h\left(\bar{\alpha}, \bar{\alpha}_t\right)-\mathbb{E}_{\gamma_t}\left[\frac{1}{\eta} V_h\left(\bar{\alpha}, \bar{\alpha}_{t+1}\right) \mid \bar{\alpha}_t\right]
\end{aligned}
\end{equation}

where the first inequality follows from Three-Point Property, the second inequality is by strongly convexity of $h$, the third inequality uses that $a^2+b^2 \ge 2ab$ for $a, b \in \mathbb{R}$ and Cauchy-Schwarz to obtain
\begin{equation}
\label{eq:util_eq25}
    \begin{aligned}
\frac{\sigma_1}{2 \eta}\left\|\bar{\alpha}_{t+1}-\bar{\alpha}_t\right\|_1^2+\frac{\eta}{\sigma_1}\left\|\tilde{g}_t\right\|_{\infty}^2 & \geq\left\|\bar{\alpha}_{t+1}-\bar{\alpha}_t\right\|_1\left\|\tilde{g}_t\right\|_{\infty} \\
& \geq\left|\left\langle\tilde{g}_t, \bar{\alpha}_t-\bar{\alpha}_{t+1}\right\rangle\right|
\end{aligned}
\end{equation}
and the last inequality follows from equation \ref{eq:util_eq24}. Taking expectation with respect to $\xi_{t-1}$ and multiplying by $\eta$ on both sides of equation \ref{eq:util_eq25} yields:
\begin{equation}
\label{eq:util_eq26}
    \begin{aligned}
& \mathbb{E}_{\xi_{t-1}}\left[\eta\left\langle g_t, \bar{\alpha}_t-\bar{\alpha}\right\rangle\right] \\
& \leq \frac{2\left(1+\bar{\rho}^2\right)}{\sigma_1} \eta^2+\mathbb{E}_{\xi_{t-1}}\left[V_h\left(\bar{\alpha}, \bar{\alpha}_t\right)\right]-\mathbb{E}_{\xi_t}\left[V_h\left(\bar{\alpha}, \bar{\alpha}_{t+1}\right)\right]
\end{aligned}
\end{equation}

Consider the process $Q_t = \sum^t_{s=1}\eta\left\langle g_s, \bar{\alpha}_s-\bar{\alpha}\right\rangle-\mathbb{E}_{\xi_{s-1}}\left[\left\langle g_s, \bar{\alpha}_s-\bar{\alpha}\right\rangle\right]$, which is martingale with respect to $\xi_{t}$ (i.e., $Q_t \in \sigma\left(\xi_t\right)$ and $\mathbb{E}\left[Q_{t+1} \mid \xi_t\right] = Q_t$) with increments bounded by
\begin{equation}
    \begin{aligned}
\left|Q_t-Q_{t-1}\right| & \leq \eta\left(\left\|g_t\right\|_{\infty}+\mathbb{E}_{\xi_{t-1}}\left\|g_t\right\|_{\infty}\right)\left\|\bar{\alpha}_t-\bar{\alpha}\right\|_1 \\
& \leq 2(1+\bar{\rho}) M\left\|\bar{\alpha}_t-\bar{\alpha}\right\|_{\infty} \\
& \leq 4 M(1+\bar{\rho})\left\|K_1\alpha^{\max }\right\|_{\infty} \\
& =4 MK_1(1+\bar{\rho})\left(\frac{\bar{f}}{\underline{\rho}}+1\right)<\infty
\end{aligned}
\end{equation}
where the first inequality is Cauchy-Schwarz, the second inequality is from $\left\|g_t\right\|_\infty \le 1 + \bar{\rho}$ almost surely, and the last inequality utilize Lemma 3. Since $\tau_A$ is a stopping time with respect to $\xi_t$ and $\tau_A$ is bounded, the Optional Stopping Theorem implies that $\mathbb{E}\left[M_{\tau_A}\right]=0$. Therefore,
\begin{equation}
\label{eq:util_eq27}
    \begin{gathered}
    \mathbb{E}\left[\sum_{t=1}^{\tau_A} \eta\left\langle g_t, \bar{\alpha}_t-\bar{\alpha}\right\rangle\right]=\mathbb{E}\left[\sum_{t=1}^{\tau_A} \mathbb{E}_{\xi_{t-1}}\left[\eta\left\langle g_t, \bar{\alpha}_t-\bar{\alpha}\right\rangle\right]\right] \\
    \leq \frac{2\left(1+\bar{\rho}^2\right)}{\sigma_1} \eta^2 \mathbb{E}\left[\tau_A\right]+V_h\left(\bar{\alpha}, \bar{\alpha}_0\right)
\end{gathered}
\end{equation}
where the inequality follows from summing up equation \ref{eq:util_eq26} from $t = 1$ to $t = \tau$, telescoping, and using that the Bregman divergence is non-negative.
On the other hand, it holds that by choosing $\bar{\alpha} = \alpha = 0$
\begin{equation}
\label{eq:util_eq28}
    \begin{aligned}
& \sum_{t=1}^{\tau_A} \eta\left\langle g_t, \bar{\alpha}_t-\bar{\alpha}\right\rangle \\
\geq & \sum_{t=1}^{\tau_A} \eta\left\langle g_t, K_2\left(\alpha_t-\alpha\right)\right\rangle \\
= & \sum_{t=1}^{\tau_A} \eta K_2\left\langle\nabla_\alpha L\left(z_t, \alpha_t\right), \alpha_t-\alpha\right\rangle \\
= & \sum_{t=1}^{\tau_A} \eta K_2\left(L\left(z_t, \alpha_t\right)-L\left(z_t, \alpha\right)\right) \\
= & \sum_{t=1}^{\tau_A} \eta K_2\left(L\left(z_t, \alpha_t\right)-P\left(z_t\right)-\alpha\left(\rho-B z_t\right)\right) \\
= & \sum_{t=1}^{\tau_A} \eta K_2\left(D\left(\alpha_t\right)-P\left(z_t\right)-\alpha\left(\rho-B z_t\right)\right) \\
\geq & \tau_A \eta K_2\left(D\left(\hat{\alpha}_{\tau_A}\right)-\frac{\sum_{t=1}^{\tau_A} P\left(z_t\right)}{\tau_A}\right)-\sum_{t=1}^{\tau_A} \alpha\left(\rho-B z_t\right) \\
= & \tau_A \eta K_2\left(D\left(\hat{\alpha}_{\tau_A}\right)-\frac{\sum_{t=1}^{\tau_A} P\left(z_t\right)}{\tau_A}\right)
\end{aligned}
\end{equation}
where the first inequality uses Lemma 2, the first equality uses equation \ref{eq:util_eq23}, the second equality is because $L\left(z,\alpha\right)$ is linear in $\alpha$, the third equality is from $z_t = \mathop{\arg \min}_zL\left(z, \alpha_t\right)$, the second inequality uses convexity of $D\left(\cdot\right)$ over $\alpha$, and the last equality is because $\alpha = 0$. Combining equation \ref{eq:util_eq27} and \ref{eq:util_eq28} and choosing $\bar{\alpha} = \alpha = 0$, we obtain:
\begin{equation}
\label{eq:util_eq29}
    \begin{aligned}
& \mathbb{E}\left[\tau_A D\left(\hat{\alpha}_{\tau_A}\right)-\sum_{t=1}^{\tau_A} P\left(z_t\right)\right] \\
& \leq \frac{2\left(1+\bar{\rho}^2\right)}{K_2\sigma_1} \eta \mathbb{E}\left[\tau_A\right]+\frac{V_h\left(0, \bar{\alpha}_0\right)}{K_2\eta}
\end{aligned}
\end{equation}

Notice that $\alpha_t$ and $z_t$ are measurable given the sigma algebra $\sigma\left(\xi_{t-1}\right)$. From the update of $x_t$ and $z_t$, we know that if a request of type $i$-th is realized in the $t$-th iteration, then $x_t = (z_t)_i/p_i$. Thus it holds for any $t \le \tau_A$ that
\begin{equation}
    \mathbb{E}_{\gamma_t}\left[f_t\left(x_t\right) \mid \xi_{t-1}\right]=\sum_{i=1}^n p_i f_i\left(\left(z_t\right)_i / p_i\right)=P\left(z_t\right)
\end{equation}

Therefore, another martingale argument yields that
\begin{equation}
\label{eq:util_eq30}
    \mathbb{E}\left[\sum_{t=1}^{\tau_A} f_t\left(x_t\right)\right]=\mathbb{E}\left[\sum_{t=1}^{\tau_A} P\left(z_t\right)\right]
\end{equation}
Combining equation \ref{eq:util_eq29} and \ref{eq:util_eq30} finishes the proof.

\subsubsection{Theorem 1}
Consider Algorithm \ref{alg: rcpacing} with step-size $\eta \leq K_1\sigma_2$ and initial dual solution $\alpha_0 \leq \alpha^{max}$. Suppose Assumption 1-5 are satisfied. Then it holds for any $T \ge 1$ that
\begin{equation}
\begin{aligned}
    \text{Regret}\left(A\right) \leq & \frac{2\left(1+\bar{\rho}^2\right)}{K_2\sigma_1} \eta T+\frac{V_h\left(0, \bar{\alpha}_0\right)}{K_2\eta} \\
& +\frac{\bar{f}}{\underline{\rho} \eta}\left\|\nabla h\left(K_1\alpha^{\max }\right)-\nabla h\left(\bar{\alpha}_0\right)\right\|_{\infty}+\frac{\overline{f}}{\underline{\rho}}.
\end{aligned}
\end{equation}

When choosing $\eta = O\left(1/\sqrt{T}\right)$, we obtain that $\text{Regret}\left(A\right) \leq O\left(\sqrt{T}\right)$ when $T$ is sufficiently large, and, therefore, our algorithm yields sublinear regret.

\noindent \textbf{Proof.}
For any $\tau_A \leq T$, we have
\begin{equation}
    \begin{aligned}
\mathrm{OPT} & =\frac{\tau_A}{T} \mathrm{OPT}+\frac{T-\tau_A}{T} \mathrm{OPT} \\
& \leq \tau_A D\left(\hat{\alpha}_{\tau_A}\right)+\left(T-\tau_A\right) \bar{f}
\end{aligned}
\end{equation}
where the inequality uses equation \ref{eq:ub_of_opt} and the fact that $\mathrm{OPT} \leq \bar{f}$ . Therefore,
\begin{equation}
\label{eq:util_eq31}
    \begin{aligned}
& \operatorname{Regret}(A) \\
= & \operatorname{OPT}-R(A) \\
\leq & \mathbb{E}_{\mathcal{P}}\left[\tau_A D\left(\hat{\alpha}_{\tau_A}\right)+\left(T-\tau_A\right) \bar{f}-\sum_{t=1}^{T} f_t\left(x_t\right)\right] \\
\leq & \mathbb{E}_{\mathcal{P}}\left[\left(\tau_A D\left(\hat{\alpha}_{\tau_A}\right)-\sum_{t=1}^{\tau_A} f_t\left(x_t\right)\right)\right] \\
& +\mathbb{E}_{\mathcal{P}}\left[\left(T-\tau_A\right) \bar{f}\right] \\
\leq & \frac{2\left(1+\bar{\rho}^2\right)}{K_2\sigma_1} \eta \mathbb{E}_{\mathcal{P}}\left[\tau_A\right]+\frac{V_h\left(0, \bar{\alpha}_0\right)}{K_2\eta} \\
& +\frac{\bar{f}}{\underline{\rho} \eta}\left\|\nabla h\left(K_1\alpha^{\max }\right)-\nabla h\left(\bar{\alpha}_0\right)\right\|_{\infty}+\frac{\overline{f}}{\underline{\rho}} \\
\leq & \frac{2\left(1+\bar{\rho}^2\right)}{K_2\sigma_1} \eta T+\frac{V_h\left(0, \bar{\alpha}_0\right)}{K_2\eta} \\
& +\frac{\bar{f}}{\underline{\rho} \eta}\left\|\nabla h\left(K_1\alpha^{\max }\right)-\nabla h\left(\bar{\alpha}_0\right)\right\|_{\infty}+\frac{\overline{f}}{\underline{\rho}}
\end{aligned}
\end{equation}
where the second inequality is because $\tau_A \leq T$ and $f_t\left(x_t\right) \geq 0$, the third inequality uses Proposition 2 and Proposition 3, and the last inequality is from $\tau_A \leq T$ almost surely. Moreover, equation \ref{eq:util_eq31} holds for any $\mathcal{P} \in \mathcal{J}$, which finishes the proof of Theorem 1.

\subsubsection{Theorem 2}
When the dual variable shifts the same distance, a higher percentile will make the participation rate of a campaign generate greater fluctuations.

\noindent \textbf{Proof.} According to Assumption 2, the impression quality (or CTR in this paper) follows a beta distribution, whose probability density function can be expressed as
\begin{equation}
    Beta\left(v|m,n\right)=\frac{1}{B\left(m,n\right)}v^{m-1}\left(1-v\right)^{n-1}
\end{equation}
\begin{equation}
B\left(m,n\right) = \int^1_0v^{m-1}\left(1-v\right)^{n-1}dv=\frac{\Gamma\left(m\right)\Gamma\left(n\right)}{\Gamma\left(m+n\right)}
\end{equation}
where $v \in [0, 1]$ denotes the impression quality, $m \ge 2$ and $n \ge 2$ are parameters of the beta distribution. 

Suppose the dual variable of a campaign is $\alpha$, then its participation rate can be obtained by 
\begin{equation}
\begin{aligned}
    \operatorname{PR}_{\alpha} &= \int^1_\alpha Beta\left(v|m,n\right)dv \\
    &= \frac{1}{B\left(m,n\right)}\int^1_\alpha v^{m-1}\left(1-v\right)^{n-1}dv
\end{aligned}
\end{equation}

When the dual variable shifts by a distance of $\delta \in (0, \alpha]$, the resulting fluctuation in the participation rate is
\begin{equation}
    \mathcal{F}_\delta\left(\alpha\right) = \frac{\operatorname{PR}_{\alpha-\delta}}{\operatorname{PR}_{\alpha}} = \frac{\int^1_{\alpha-\delta} v^{m-1}\left(1-v\right)^{n-1}dv}{\int^1_\alpha v^{m-1}\left(1-v\right)^{n-1}dv}
\end{equation}

Let $g\left(\alpha\right)=\alpha^{m-1}\left(1-\alpha\right)^{n-1}$ and $h\left(\alpha\right)=\int^1_\alpha g\left(v\right)dv$, the original problem can be formulated as the task of demonstrating $\mathcal{F}_\delta\left(\alpha\right)$ is monotonically increasing when $0 < \delta \le \alpha < 1$. The derivative of $\mathcal{F}_\delta\left(\alpha\right)$ is
\begin{equation}
\begin{aligned}
    \mathcal{F}'_\delta\left(\alpha\right) &= \frac{g\left(\alpha\right)h\left(\alpha-\delta\right)-g\left(\alpha-\delta\right)h\left(\alpha\right)}{h^2\left(\alpha\right)} \\
    &= \frac{1}{h^3\left(\alpha\right)h\left(\alpha-\delta\right)}\cdot\left[\frac{g\left(\alpha\right)}{h\left(\alpha\right)}-\frac{g\left(\alpha-\delta\right)}{h\left(\alpha-\delta\right)}\right]
\end{aligned}
\end{equation}

Since $h\left(\alpha\right) > 0$ when $0 \leq \alpha < 1$, the problem can be converted to demonstrating that $\phi\left(\alpha\right)=\frac{g\left(\alpha\right)}{h\left(\alpha\right)}$ is monotonically increasing. The derivative of $\phi\left(\alpha\right)$ is
\begin{equation}
\begin{aligned}
    \phi'\left(\alpha\right) &= \frac{g'\left(\alpha\right)h\left(\alpha\right)-g\left(\alpha\right)h'\left(\alpha\right)}{h^2\left(\alpha\right)} \\
    &= \frac{g'\left(\alpha\right)h\left(\alpha\right)+g^2\left(\alpha\right)}{h^2\left(\alpha\right)}
\end{aligned}
\end{equation}

We only need to prove that $\Phi\left(\alpha\right)=g'\left(\alpha\right)h\left(\alpha\right)+g^2\left(\alpha\right) \ge 0$ when $0 \leq \alpha < 1$. Let
\begin{equation}
\begin{aligned}
    & a_i=\alpha+\frac{1-\alpha}{k}, \; b_i=\alpha+\frac{1-\alpha}{k}i \\
    & c_i=\alpha, \; d_i=\alpha+\frac{1-\alpha}{k}i+\frac{1-\alpha}{k}
\end{aligned}
\end{equation}
where $i=1,2,\cdots,n-1$. We have
\begin{equation}
    a_i+b_i=c_i+d_i=2\alpha+\frac{1-\alpha}{k}(i+1)=\mathcal{M}_i
\end{equation}
where $0 \le c_i \le a_i \le \mathcal{M}_i/2 \le b_i \le d_i \le \mathcal{M}_i$.

Let $q_i\left(\alpha\right) = \alpha\left(\mathcal{M}_i-\alpha\right)$, since $q\left(\alpha\right)$ is monotonically increasing on $\left[0, \mathcal{M}_i/2\right]$ and $0 \le c_i \le a_i \le \mathcal{M}_i/2$, we have $a_ib_i=q_i\left(a_i\right) \ge q_i\left(c_i\right) = c_id_i$, and
\begin{equation}
\label{eq:util_eq83}
\begin{aligned}
    &g\left(\alpha+\frac{1-\alpha}{k}\right)g\left(\alpha+\frac{1-\alpha}{k}i\right) \\
    &\quad-g\left(\alpha\right)g\left(\alpha+\frac{1-\alpha}{k}+\frac{1-\alpha}{k}i\right)\\
    &=g\left(a_i\right)g\left(b_i\right)-g\left(c_i\right)g\left(d_i\right) \\
&= \left(a_ib_i\right)^{m-1} \cdot \left(1-\mathcal{M}_i+a_ib_i\right)^{n-1} \\
&\quad-\left(c_id_i\right)^{m-1} \cdot \left(1-\mathcal{M}_i+c_id_i\right)^{n-1} \ge 0
\end{aligned}
\end{equation}

From the definition of derivative and integration, we can obtain
\begin{equation}
\begin{aligned}
&\Phi\left(\alpha\right) =g'\left(\alpha\right)h\left(\alpha\right)+g^2\left(\alpha\right) \\
&= \lim_{k \rightarrow \infty}\frac{g\left(\alpha+\frac{1-\alpha}{k}\right)-g\left(\alpha\right)}{\frac{1-\alpha}{k}}\sum^k_{i=0}\frac{1-\alpha}{k}g\left(\alpha+\frac{1-\alpha}{k}i\right) \\
& \quad + g^2\left(\alpha\right) \\
&= \lim_{k \rightarrow \infty}\left[g\left(\alpha+\frac{1-\alpha}{k}\right)-g\left(\alpha\right)\right]\sum^k_{i=0}g\left(\alpha+\frac{1-\alpha}{k}i\right) \\
& \quad + g^2\left(\alpha\right) \\
&= \lim_{k\rightarrow \infty}\sum^k_{i=1}g\left(\alpha+\frac{1-\alpha}{k}\right)g\left(\alpha+\frac{1-\alpha}{k}i\right)- \\
& \quad \lim_{k\rightarrow \infty}\sum^{k-1}_{i=1}g\left(\alpha\right)g\left(\alpha+\frac{1-\alpha}{k}+\frac{1-\alpha}{k}i\right) \\
&= \lim_{k\rightarrow \infty}\sum^{k-1}_{i=1}{\left[g\left(\alpha+\frac{1-\alpha}{k}\right)g\left(\alpha+\frac{1-\alpha}{k}i\right)\right.} \\
& \quad \left.-g\left(\alpha\right)g\left(\alpha+\frac{1-\alpha}{k}+\frac{1-\alpha}{k}i\right)\right] \\
& \quad +\lim_{k\rightarrow \infty}g\left(\alpha+\frac{1-\alpha}{k}\right)g\left(1\right) \\
&= \lim_{k \rightarrow \infty}\sum^{k-1}_{i=1}{\left[g\left(\alpha+\frac{1-\alpha}{k}\right)g\left(\alpha+\frac{1-\alpha}{k}i\right)\right.} \\
& \quad \left. -g\left(\alpha\right)g\left(\alpha+\frac{1-\alpha}{k}+\frac{1-\alpha}{k}i\right)\right] \geq 0,
\end{aligned}
\end{equation}
where the last equality is because $g\left(1\right)=0$, and the last inequality uses equation \ref{eq:util_eq83}. This finishes the proof of Theorem 2.

\subsubsection{Theorem 3}
When the distribution of impression quality drifts (assume $m$ and $n$ do not change simultaneously), a higher percentile will make the participation rate of a campaign generate greater fluctuations.

\noindent \textbf{Proof.} Denote the participation rate of a campaign as
\begin{equation}
\begin{aligned}
    \operatorname{PR}_{\alpha}\left(m, n\right) &= \int^1_\alpha Beta\left(v|m,n\right)dv \\
    &= \frac{\int^1_\alpha v^{m-1}\left(1-v\right)^{n-1}dv}{\int^1_0 v^{m-1}\left(1-v\right)^{n-1}dv}
\end{aligned}
\end{equation}

To demonstrate its monotonicity with regard to $m$, the partial derivative of $\operatorname{PR}_{\alpha}\left(m, n\right)$ with respect to $m$ is deduced as
\begin{equation}
\label{eq:util_eq86}
    \frac{\partial PR_\alpha}{\partial m}=\frac{h\left(0\right) \cdot \hat{h}\left(\alpha\right)-\hat{h}\left(0\right) \cdot h\left(\alpha\right)}{h^2\left(0\right)}
\end{equation}
where $h\left(\alpha\right)=\int^1_\alpha v^{m-1}\left(1-v\right)^{n-1}dv$ and $\hat{h}\left(\alpha\right)=\int^1_\alpha \ln v\cdot v^{m-1}\left(1-v\right)^{n-1}dv$. Let $\mathcal{H}\left(\alpha\right)=h\left(0\right) \cdot \hat{h}\left(\alpha\right)-\hat{h}\left(0\right) \cdot h\left(\alpha\right)$, its derivative with respect to $\alpha$ is
\begin{equation}
    \mathcal{H}'\left(\alpha\right)=\alpha^{m-1}\left(1-\alpha\right)^{n-1}\left[\hat{h}\left(0\right)-h\left(0\right)\ln \alpha\right]
\end{equation}

It is obvious that $\mathcal{H}'\left(\alpha\right)$ is positive on $\left[0, \alpha_0\right]$ and negative on $\left[\alpha_0,1\right]$, where $\alpha_0=\hat{h}\left(0\right)/h\left(0\right)$. Thus, $\mathcal{H}\left(\alpha\right)$ increases on $\left[0, \alpha_0\right]$ and decreases on $\left[\alpha_0,1\right]$. Since $\mathcal{H}\left(0\right)=\mathcal{H}\left(1\right)=0$, it can be deduced that $\mathcal{H}\left(\alpha\right) \ge 0$ when $\alpha \in \left[0,1\right]$. Together with equation \ref{eq:util_eq86}, it can be proved that $\operatorname{PR}_{\alpha}\left(m, n\right)$ increases monotonically with respect to $m$. Similarly, we can prove that $\operatorname{PR}_{\alpha}\left(m, n\right)$ decreases monotonically with respect to $n$.

When $m$ has an increase of $\delta$, the resulting fluctuation in the participation rate is
\begin{equation}
\begin{aligned}
\mathcal{F}_\delta\left(\alpha\right) &= \frac{\operatorname{PR}_{\alpha}\left(m+\delta, n\right)}{\operatorname{PR}_{\alpha}\left(m, n\right)} \\
&=V\left(m,n,\delta\right)\cdot\frac{\int^1_\alpha v^{m+\delta-1}\left(1-v\right)^{n-1}dv}{\int^1_\alpha v^{m-1}\left(1-v\right)^{n-1}dv}
\end{aligned}
\end{equation}

Our target is to prove that $\mathcal{F}_\delta\left(\alpha\right)$ is monotonically increasing with respect to $\alpha$ for any $\alpha \in [0, 1)$. The derivative of $\mathcal{F}_\delta\left(\alpha\right)$ is obtained by
\begin{equation}
\begin{aligned}
&\mathcal{F}_\delta'\left(\alpha\right) \\
&= \frac{\alpha^{m-1}\left(1-\alpha\right)^{n-1}\int^1_\alpha\left(v^\delta-\alpha^\delta\right)v^{m-1}\left(1-v\right)^{n-1}dv}{\left[\int^1_\alpha v^{m-1}\left(1-v\right)^{n-1}dv\right]^2} \\
&\ge 0
\end{aligned}
\end{equation}
which indicates that a greater $\alpha$ will result in a more pronounced fluctuation in the participation rate as $m$ varies. Through a similar process, we can also prove that when $n$ changes, a greater $\alpha$ will result in a more pronounced fluctuation in the participation rate, which is omitted here.
\clearpage

\end{document}